\documentclass[10pt]{article}

\usepackage{amsmath}
\usepackage{amssymb}

\usepackage{graphicx}

\usepackage{cite}

\usepackage{color} 

\usepackage[labelfont=bf,labelsep=period,justification=raggedright]{caption}

\usepackage{amssymb,amsfonts,amsmath}
\usepackage{latexsym}
\usepackage{epsfig}
\usepackage{url}
\usepackage{microtype}
\usepackage{fixmath}
\usepackage{bm}

\newcommand{\bx}{\bm{x}}

\newcommand{\bz}{\bm{z}}
\newcommand{\Z}{\bm{Z}}
\newcommand{\1}{\bm{1}}
\newcommand{\comment}[1]{}

\makeatletter
\renewcommand{\@biblabel}[1]{\quad#1.}
\makeatother

\date{}

\begin{document}

\begin{flushleft}
{\Large
\textbf{Posterior predictive checks to quantify lack-of-fit in
  admixture models of latent population structure}
}
\\
David Mimno$^{1}$, 
David M Blei$^{2}$, 
Barbara E Engelhardt$^{3,\ast}$
\\
\bf{1} Department of Information Science, Cornell University
\\
\bf{2} Departments of Statistics and Computer Science, Columbia University
\\
\bf{3} Departments of Biostatistics \& Bioinformatics and Statistical Science, Duke University
\\
$\ast$ E-mail: barbara.engelhardt@duke.edu
\end{flushleft}

\section*{Abstract}
Admixture models are a ubiquitous approach to capture latent population structure in genetic samples. Despite the
  widespread application of admixture models, little thought has been
  devoted to the quality of the model fit or the accuracy of the
  estimates of parameters of interest for a particular study. Here we
  develop methods for validating admixture models based on posterior
  predictive checks (PPCs), a Bayesian method for assessing the
  quality of a statistical model.  
  We develop PPCs for five
  population-level statistics of interest: within-population genetic variation,
  background linkage disequilibrium, number of ancestral populations, between-population genetic variation, and the downstream use of admixture parameters to
  correct for population structure in association studies. Using PPCs,
  we evaluate the quality of the model estimates for four
  qualitatively different population genetic data sets: the POPRES
  European individuals, the HapMap phase 3 individuals, continental
  Indians, and African American individuals.  We found that the same
  model fitted to different genomic studies resulted in highly
  study-specific results when evaluated using PPCs, illustrating the
  utility of PPCs for model-based analyses in large genomic studies.

\section{Introduction}

One of the essential problems for population genetics is to
characterize latent population structure in genetic samples.  Inferred
population structure is used to control for confounding effects in
both genome-wide association studies (GWAS) and quantitative trait
mapping~\cite{Pritchard2000b,Price2006}, and to explore genetic
relationships when studying population ancestry and
history~\cite{reich2011denisova,moorjani2011history,wang2013apparent}.



An important and influential approach to characterizing latent
population structure is the \emph{admixture model}, first implemented
in the \textsc{structure} program~\cite{pritchard2000inference}. The
admixture model is a Bayesian model of a collection of genomes.  It
represents each genome as a convex combination of $K$ ancestral
populations, and describes each ancestral population by
population-specific allele frequencies across every genetic locus.
Given observed genetic data, admixture models recover both the genetic
variation within ancestral populations and the proportions of each
ancestral population within each genome.  Because of their descriptive
power, admixture models have become essential for exploratory analyses
of genomic studies~\cite{Gilbert2012}; they have transformed modern
research in population genetics.

The admixture model, like all statistical models, makes assumptions to
simplify the data. Through these assumptions, it models complex
genomic data in a way that is both analytically useful and
computationally tractable.  For example, the original admixture model
assumes that individuals in the sample are distantly related, that all
population- and locus-specific allele frequencies are equally likely,
and that genetic loci are independent.  Population genetics tells us,
however, most genomic data violate these
assumptions~\cite{pritchard2000inference}.  Paraphrasing the famous
quip by statistician George Box, our question is not whether the
model is true---we know that it is not---but whether it is useful.
Do the fitted model parameters help with the analytic task, or do the
model's simplifying assumptions lead scientists astray to unsupported
conclusions?

In this paper, we show that the effect of the admixture assumptions on
inference of latent population structure depends largely on the data at hand; thus diagnosing
model misspecification should become a regular practice in the
application of admixture models.  Without checking the model when
applied to a given data set, scientists may find spurious associations
between diseases and genetic variants while believing they are
correcting for latent structure, or follow blind alleys of ancestral
history while exploring inferred population structure that is only an
artifact of the admixture's simplifying assumptions.  When we fit the
admixture model to genetic data---whether we are exploring latent
structure or using inferred population structure in downstream
analyses---we rely on the recovered representation from a statistical
procedure to be meaningfully connected to the true genetic structure
that has emerged from a complex evolutionary process.  It is essential
that we assess the strength of this assumed connection.

To this end, we develop a family of statistical procedures for
checking the goodness-of-fit of the admixture model to genomic data.
Our procedures are based on \textit{posterior predictive checks}
(PPCs), a technique from Bayesian statistics that is used to quantify
the effect of Bayesian model
misspecification~\cite{Box1980,Rubin1984,Meng1993,gelman1996posterior,Gelman2013}.
A PPC works as follows~\cite{Rubin1984}.  We first fit a model using
the observed data, estimating the posterior distribution of the latent
parameters.  The fitted model induces a distribution of future data
conditioned on the observations; this distribution is called the
\textit{posterior predictive distribution}.  We next use the posterior
predictive distribution to generate several synthetic data sets.
Finally, we check whether the simulated data sets are close to the
observed data set when summarized through a statistic of interest,
called the \emph{discrepancy function}.  The idea is that if the model
assumptions are appropriate then data generated from the posterior
predictive distribution will look like the observed data, and the
discrepancy measures a relevant property of the data that we hope to
represent.  If the model is well specified for a specific data set,
then the observed data, viewed through the discrepancy, will be a
likely draw from the estimated posterior predictive distribution.  If
the model is not well specified then the observed discrepancy will
look like an outlier.



\begin{figure}
\vspace{-10mm}
\begin{center}
\includegraphics[height=3in]{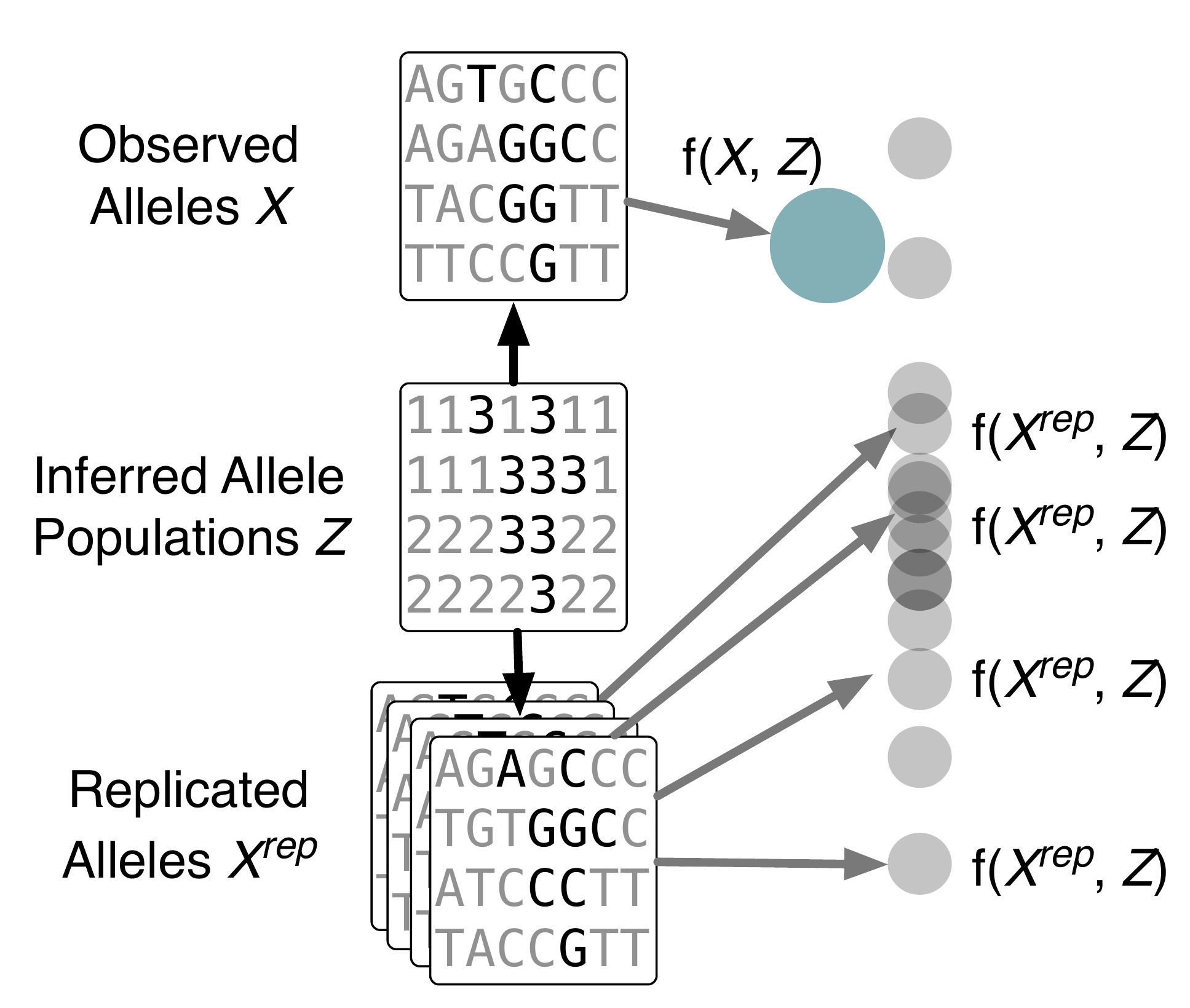}
\caption{{\bf Schematic diagram showing posterior predictive checks
    (PPCs) applied to genetic studies.} The admixture model is
  fitted to a collection of genomic data. A distribution of the
  discrepancy function $f(\cdot)$ for each of $K$ inferred populations to the replicated genomic data $X^{rep}$, given the estimated ancestral population parameters $Z$, approximates the
  posterior predictive distribution. The probability of
  observing the discrepancy function applied to the observed data $X$ with
  respect to this approximate posterior predictive distribution
  quantifies the goodness-of-fit.}
\label{fig:ppc-schematic}
\end{center}
\end{figure}


Given observed genotype data, checking admixture models with a PPC
works as follows (Figure~\ref{fig:ppc-schematic}). We first fit an
admixture model to the data, estimating ancestral populations and
individual-specific population proportions.  Note that most analyses
end here, e.g., with illustrations of the population proportions as in
Figure~\ref{rainbow-diagram}.  We then simulate genomes from the
posterior predictive distribution, using posterior estimates of the
latent parameters to draw synthetic genetic data that share the same
latent population structure as the observed data; we repeat this
process many times to create a collection of \emph{replicated data
  sets}.  Finally, we evaluate and compare discrepancy functions on
both the replicated data sets and the observed data.  The discrepancy
may be a function of both observed and latent
variables~\cite{Meng1993,gelman1996posterior}, and our discrepancies
measure important population statistics including within-population genetic variance and background linkage disequilibrium (LD).  Specifically, we compare discrepancies computed on the
observed data to the empirical distribution of the discrepancies
computed on the replicated data.  When an observed discrepancy is not
likely relative to the replicated discrepancies, the PPC suggests that
the model is misspecified (with respect to the discrepancy) for the
observed data.  We maintain that PPC assessments are best made
visually~\cite{gelman1996posterior,Gelman2013}.



\begin{figure*} \begin{center}
    \includegraphics[width=7.1in,height=2.7in]{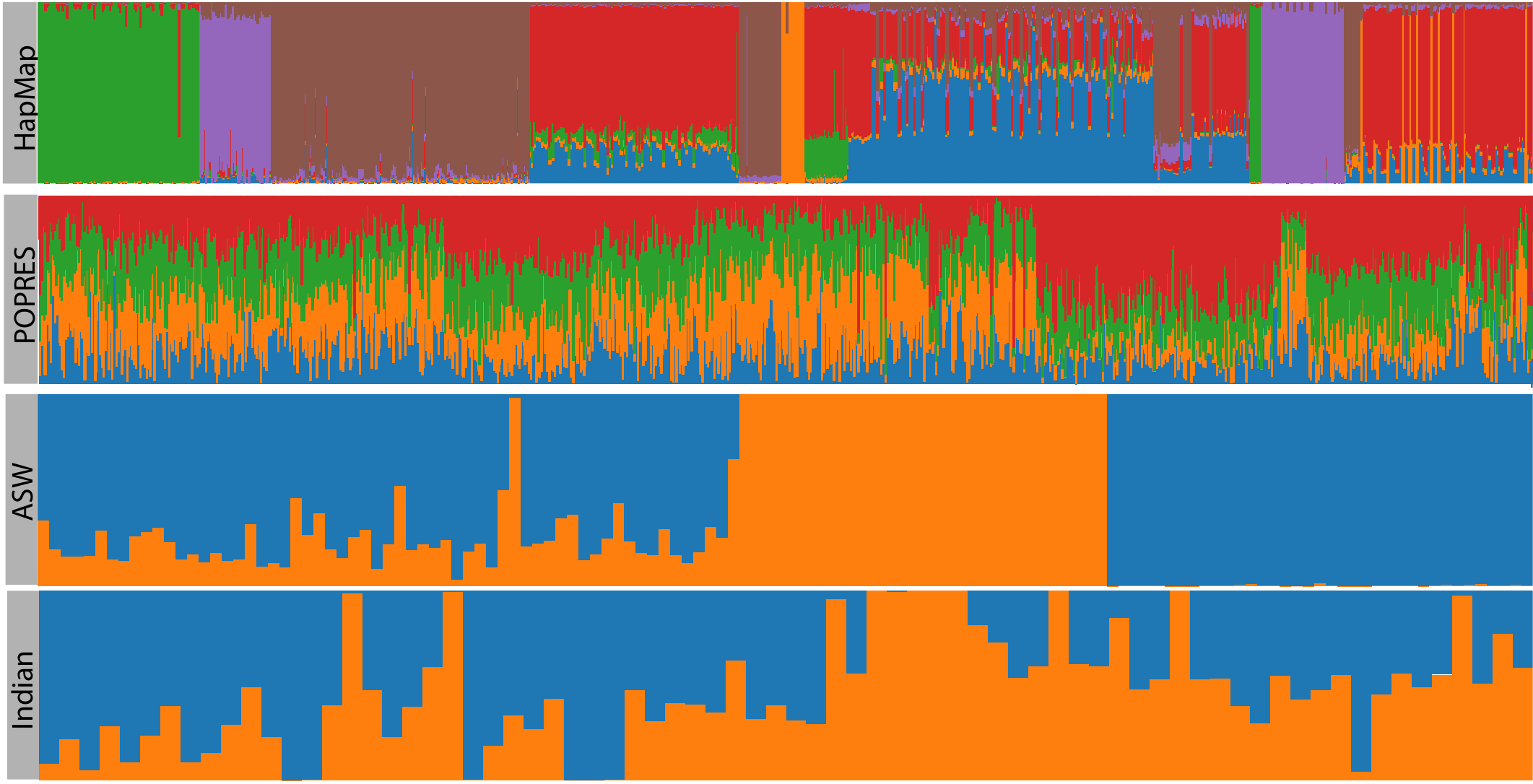}
    \caption{{\bf Illustration of the variation in individual-specific
        ancestry proportions across the four genomic studies.} The
      $x$-axis represents the individuals in each study, and the
      $y$-axis is the proportion of the genome with maximum likelihood
      assignment in each fitted ancestral populations (each has its
      own color). Panel A: HapMap phase 3 individuals, clustered by geographic
      origin of sample, fitted to six ancestral
      populations~\cite{Rosenberg2002}. Individuals in this study, for
      the most part, have ancestry in one of the six continental
      ancestral populations, and are not substantially admixed. Panel
      B: POPRES individuals, clustered by geographic location in
      Europe, fitted with four ancestral
      populations~\cite{Engelhardt2010, Nelson2008}. Because of the
      genetic proximity of the four ancestral populations,
      representing four corners of Europe, each individual has some
      proportion of ancestry in each of the populations. Panel C: ASW
      individuals, clustered by reported African ancestry, fitted to
      two ancestral populations~\cite{1000genomes}. Because we include
      African and European individuals in this sample, we see
      individuals have either approximately $20\%$ African and $80\%$
      European ancestry, or all African or all European
      ancestry. Panel D: Indian individuals, fitted to two ancestral
      populations~\cite{Reich2009}. Most individuals here have some
      proportion of ancestry in each of the two ancestral populations,
      because of an ancient admixture event.}
\label{fig:rainbow-diagram}
\end{center}
\end{figure*}

We used PPCs to check for misspecification in four genomic data sets:
HapMap phase 3, Europeans, African Americans, and continental
Indians. These data have been previously characterized using an
admixture model and have qualitatively different types of latent
population structure (Figure~\ref{fig:rainbow-diagram}). We developed
five discrepancy functions to check for important types of model
misspecification in common admixture model analyses. We based these
discrepancy functions on common measures in population genetics:
\begin{itemize}

\item \emph{Identity by state (IBS)}: we test for the impact of long-range SNP
  correlations on within-population variance estimates by quantifying
  the genomic variation among pairs of individuals within alleles from the same
  ancestral population;

\item \emph{Background LD}: we test for the impact of short-range SNP
  correlations on population-specific allele frequency estimates by
  computing the autocorrelation between SNPs, or background LD;

\item \emph{$F_{ST}$}: we test the appropriate numbers of ancestral
  populations by computing $F_{ST}$ among labelled and inferred
  ancestry;

\item \emph{Assignment uncertainty}: we test how distinct the
  ancestral populations are from one another by quantifying
  uncertainty in ancestral population assignment;

\item \emph{Association tests}: we test whether or not the inferred
  population structure adequately controls for confounding latent population
  structure in association mapping studies by quantifying the
  difference in statistical significance of corrected associations
  versus uncorrected associations under the null hypothesis of no
  association.
\end{itemize}
As we note in the Discussion, using and comparing PPCs with
several discrepancies lets us innovate the model in the directions for
which it is most important.

With five discrepancies and four data sets, PPCs reveal that each
application of the admixture model meets and diverges from its
assumptions differently. Each PPC indicates when we might extend the
model to better match the complexities of the study data, and in the
Discussion we discuss specific extensions to address each model
application for which we find the model is misspecified.  While we
focus here on the admixture model, we emphasize that assessing the
model fitness is important in any application of statistical models to
genetic data.  PPCs give a framework for visually and quantitatively
understanding how and when inferred latent variables cannot be safely
interpreted or are unreliable for use in downstream analysis.


\section{Results}

We used PPCs to assess the fit of an admixture model to data from four
genomic studies. We developed five \emph{discrepancy functions}, each
of which measures the degree to which the fitted model captures one
aspect of the data.  These discrepancies are functions of data and
latent structure, and are based on common statistics in population
genetics that are important to capture when using or interpreting
inferred latent variables.  They are (i) inter-individual similarity
(ii) linkage disequilibrium within genomes (iii) similarity between
inferred ancestral populations and reported geographic labels (iv)
distances between ancestral populations and (v) use of model
parameters to control for latent population structure in downstream
association studies.  Here we report our findings.

We first describe the admixture model and outline our procedure for
using a PPC to check for misspecification. The \emph{admixture model}
uses allele frequencies across genomes to recover individual-specific
ancestry proportions and population-specific allele frequencies. The
observed data include $i = 1,\dots,n$ individuals, each with $\ell \in
1,\dots,L$ single nucleotide polymorphisms (SNPs). Each SNP $\ell$ for
individual $i$ is represented as two binary variables, $x_{i,\ell,1},
x_{i,\ell,2} \in \{0,1\}$, where $x_{i,\ell,1}+x_{i,\ell,2}$ is the
number of copies of the minor (less frequent) \emph{allele} (genetic
variant). The admixture model assumes that there are $k = 1,\dots,K$
ancestral populations that describe these data, where $K$ is specified
\emph{a priori}, and that each population is associated with
location-specific allele frequencies $\beta_{k, \ell}$.  It assumes
that individual $i$'s genotype data are generated as follows: a) draw
individual-specific ancestry proportions $\theta_i$ from a uniform
Dirichlet distribution (i.e., parameter $\alpha = 1$); b) for each
SNP $\ell$, draw two categorical variables $z_{i,\ell,1}$ and
$z_{i,\ell,2}$ from a multinomial distribution with parameter
$\theta_i$; these latent variables indicate the ancestral populations
assigned to the two copies of that SNP; c) conditioned on the assigned
ancestral populations, draw two alleles for SNP $\ell$ using the
corresponding allele frequencies; that is, draw each $x_{i,\ell,\cdot}$ from a Bernoulli
with parameter $\beta_{z_{i,\ell,\cdot},\ell}$.

Conditioned on data, the admixture model estimates the posterior
distribution of the latent population structure.  This structure is
encoded in the ancestral population proportions $\theta_i$
(illustrated in Figure~2), the population-specific allele frequencies
$\beta_k$, and the assigned ancestral populations $z$.  A sampling approach
for estimating this posterior is implemented in the \emph{structure}
software~\cite{pritchard2000inference}; here we use expectation maximization (EM).  
See Methods for the complete admixture model specification
and a description of our method for parameter estimation.

Interpreting and using the posterior probabilities of ancestral proportions
assumes that the model is a good
fit to the data; the PPCs we developed check this fit.  For each
observed genomic data set $x$ (and number of populations $K$), we
first estimated the latent parameters: $\beta$, $\theta$, and $z$.
With these estimates, we generated a collection of replicated
genotype data sets.  For each one, we drew $L$ SNPs for the same $n$
individuals, holding the latent population structure $z$ and $\beta$
fixed.  This results in a set of simulated individuals $x^{rep}$.  In
particular, for all $i$ and $\ell$, we drew $x^{rep}_{i,\ell,1} |
z_{i,\ell,1}, \beta_{\ell} \sim Bernoulli(\beta_{z_{i,\ell,1},\ell})$
and similarly for $x^{rep}_{i,\ell,2}$.  Note that the replicated data
were generated conditional on the inferred latent variables; we do not
need to re-estimate them at any point in our analysis.  For each data
set $x$ and fitted admixture model we generated 100 replicated data sets $x^{rep}$ (see
Methods).



For each PPC, we
developed a discrepancy function $f(x, z, \theta)$, which is a function of
the data and latent structure.  In our PPCs, each discrepancy 
function partitions the alleles by assigned ancestral population and produces 
$K$ scalar values.   We computed
the \emph{observed discrepancy} $f(x, z, \theta)$ and the
\emph{replicated discrepancy} $f(x^{rep}, z, \theta)$ for each
replicated data set.  The empirical distribution of $f(x^{rep}, z, \theta)$ is an estimate of the
posterior predictive distribution of the discrepancy.  Thus we
checked model fitness by locating the observed discrepancy in this
distribution.  If we found that the observed discrepancy was an outlier with respect to this estimate
of the posterior predictive distribution, then we concluded that the model is not a
good fit to our data with respect to the discrepancy.

For each PPC, we used statistical visualizations and assessments of
significance to summarize the results~\cite{Gelman2004}.  The PPC plots visualize the observed
discrepancy against its posterior predictive distribution; model
misspecification is indicated by observed discrepancies that appear to be
outliers. We plotted
the value of the replicated discrepancies $f(x^{rep}, z=k, \theta)$ with
grey circles and the observed discrepancy $f(x, z=k, \theta)$
with a offset solid circle. We colored the observed discrepancy to code its
\emph{$z$-score}, the number of standard deviations from the mean of
the replicated discrepancy.   Finally, we calculated the likelihood ratio that the
$z$-scores were jointly generated from a standard normal distribution. The number of grey stars at the top of each figure corresponds to
the level of deviation from standard normal (see Methods), and these stars quantifies the magnitude of
model misspecification with respect to a discrepancy.

Our results include evaluations of four genomic studies (see Methods
for preprocessing details). For simplicity, we set the number of
ancestral populations from the research literature.  (We extend the
results to many $K$ in Supplemental Information.)  We study the following data sets:
\begin{itemize}
\item The HapMap phase 3 (\emph{HapMap}) data include $1,043$ individuals genotyped at $468,167$ SNPs from a number of populations worldwide; we expect these individuals will come from distinct (and, in a few cases, admixtures of distinct) ancestral populations~\cite{Altshuler2010}. Each individual is labeled with the location of the sample collection. Following prior work, we set $K=6$~\cite{Rosenberg2003}.
\item The POPRes (\emph{POPRES}) data include $1,387$ individuals genotyped at $197,146$ SNPs from a continental Europe~\cite{Nelson2008,Novembre2008}. We expect the ancestry of these individuals to be an admixture of populations defined on a continuous geographic gradient instead of distinct ancestral populations~\cite{Engelhardt2010}. Each individual has a label describing the geographic birthplace of their grandparents, and we included individuals only if all four grandparents shared the same birthplace. Following prior work, we set $K=4$~\cite{Engelhardt2010}.
\item The African American (\emph{ASW}) data include $61$ African American individuals from the Southwest US, $32$ individuals from Utah of Northern and Western European ancestry (CEU), and $37$ Yoruba individuals from Ibadan, Nigeria (YRI) included in the 1000 Genomes Project genotyped and sequenced at $88,885$ SNPs~\cite{1000genomes}. We expect the ancestry of the genotypes from the ASW individuals are admixed from two distinct and distant ancestral populations: European and Yoruban (West African). Individuals are labeled as ASW, CEU or YRI. We set $K=2$ because all individuals are thought to have ancestry from only European and Yoruban populations.
\item The Indian data (\emph{Indian}) include $74$ individuals genotyped at $196,375$ SNPs from $15$ tribal groups across India~\cite{Reich2009}. The ancestry of each of these individuals' genotypes is admixed between two distinct but closely related ancestral groups: North Ancestral Indian and South Ancestral Indian, and there are tribal labels associated with each sample. Following prior work, we set $K=2$~\cite{Reich2009}.
\end{itemize}
We number and describe the estimated ancestral populations in Table~\ref{tab:pops_defined}; see also Figure~2. We are using a fixed number of ancestral populations for each study, and we keep this numbering consistent throughout the figures presented in the Results.

Below, we describe each discrepancy and how each study fared under its
lens. Then, we shift the focus to the genomic studies, summarizing how
well the model fits the data across the collection of discrepancies.


\begin{table}[htdp]
\caption{{\bf Numbering and composition of each of the estimated ancestral populations for each study.} Proportions in parentheses are the estimated proportions for individuals with those geographic labels that have ancestry in that estimated ancestral population. ASW = African Americans from the Southwest of the USA; YRI = Yoruban; CEU = European.}
\begin{center}
\begin{tabular}{c|l}\hline
 \multicolumn{2}{c}{HapMap}\\
\hline
1 & Pakistan (0.6); N Europe (0.2); Russia (0.2); Israel (0.2)  \\ 
2 & New Guinea (1.0); S Europe (0.2)  \\ 
3 & S Africa (1.0); C Africa (1.0); N Africa (0.3)  \\ 
4 & N Europe (0.7); Israel (0.7); N Africa (0.7); S Europe (0.7)  \\ 
5 & S America (1.0); C America (0.9)  \\ 
6 & Japan (0.9); China (0.9); SE Asia (0.8); Russia (0.3)  \\
\hline
 \multicolumn{2}{c}{POPRES}\\
\hline
1 & Ukraine (0.5); Slovenia (0.5); Croatia (0.4); Bosn. Herz. (0.4)  \\ 
2 & Norway (0.6); Latvia (0.5); Finland (0.5); Denmark (0.5)  \\ 
3 & Portugal (0.3); Spain (0.3); Bulgaria (0.3); Switz. (0.3)  \\ 
4 & Slovakia (0.6); Turkey (0.6); Greece (0.5); Italy (0.5)  \\ \hline
 \multicolumn{2}{c}{ASW}\\
\hline
1 & YRI (1.0); ASW (0.8)  \\ 
2 & CEU (1.0); ASW (0.2)\\
 \hline
 \multicolumn{2}{c}{Indian}\\
\hline
1 & Kurumba (0.8); Bhil (0.8); Tharu (0.7); Vaish (0.7)  \\ 
2 & Vysya (1.0); Velama (0.8); Srivastava (0.8); Kamsali (0.7)  \\
 \hline
\end{tabular}
\end{center}
\label{tab:pops_defined}
\vspace{-5mm}

\end{table}%

\vspace{-5mm}
\subsection{Discrepancy in within-population genomic variation}

Population admixture models are often used to explain similarities
between individuals' genotypes: if two individuals have portions of
their chromosomes that are descended from the same ancestral
population, they will, probabilistically, have more alleles in common
than two individuals descended from distinct ancestral
populations according to the admixture model. How many more alleles these individuals have in common
depends on the separation between ancestral populations, the
proportion of the genomes with shared ancestry, and within-population
genetic variation. 

We developed a discrepancy function to compute the \emph{identity by
  state} (IBS) between pairs of individuals, which measures
conditional similarity of alleles in unrelated
individuals~\cite{Bishop1990}. For a pair of individuals, we averaged
the number of alleles in common across the genome for SNPs that share
the same ancestry assignment in the fitted model~\cite{Weir1984}. For
this discrepancy function, a pair of individuals with a discrepancy of
$1.0$ for population $k$ indicates that $100\%$ of the alleles
assigned to population $k$ in both individuals are identical and $0.0$
indicates $100\%$ different alleles. A misspecified model with respect to a
discrepancy function measuring within-population variation may imply that we cannot safely ignore
\emph{admixture LD}, or (possibly long distance) genetic correlations
induced by recent admixture of distinct populations, in our admixture
model~\cite{Stephens1994}.

\begin{figure}[t]
\begin{center}
\includegraphics[scale=0.6]{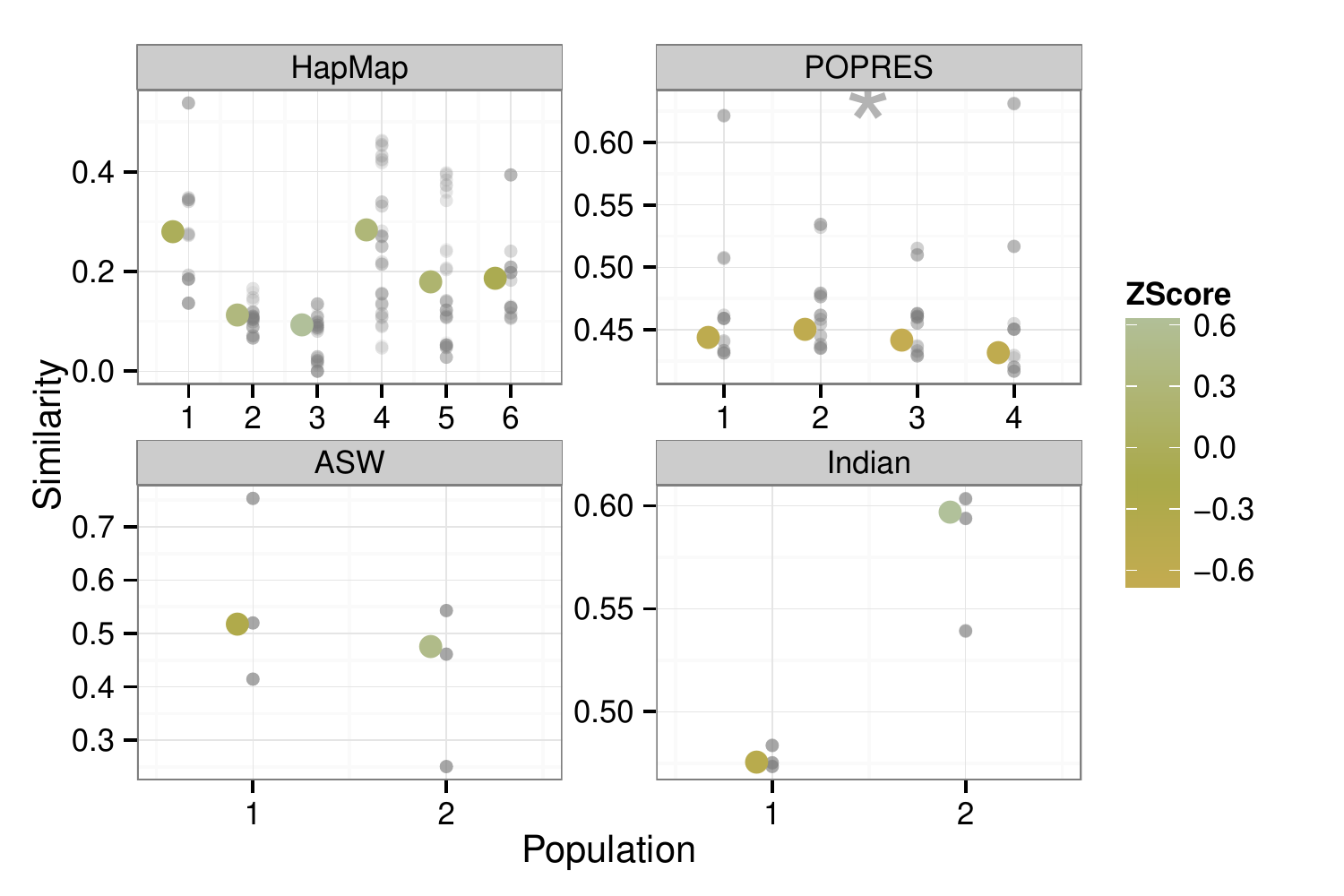}
\caption{{\bf Average population-specific inter-individual similarity
    across four studies.} Each $x$-axis represents the study for which
  the admixture model was fitted. The $y$-axis represents the mean
  inter-individual similarity across individuals conditioned on each
  ancestral population. Small semi-transparent points represent values
  from replicated data. Larger points represent values from real data,
  colored by their $z$-scores relative to the empirical distributions
  of the replicated values. Stars indicate significant divergence in
  $z$-scores from a standard normal. POPRES data show consistently
  underestimated average inter-individual similarity relative to the
  data replicates.}
\label{fig:gsk-hapmap-sim}
\end{center}
\vspace{-3mm}
\end{figure}

We performed the PPC and found that HapMap, ASW, and Indian
results show that the within-population variation is well captured
by the admixture model (Figure~\ref{fig:gsk-hapmap-sim}). In contrast, the POPRES data show consistently
underestimated average inter-individual similarity, which indicates
model misspecification for this discrepancy. Note that the two studies
with distinct, well-separated populations (HapMap, ASW) tended to have
$z$-scores greater than zero indicating greater than expected
inter-individual similarity on the observed data; the two studies with
less well-separated populations (POPRES, Indian) have $z$-scores below
zero, indicating less similarity than expected.

We also looked at the observed discrepancy without the context of the replicated discrepancy.
In the POPRES data and,
to a lesser degree, the Indian data, the average similarity of
individuals is constant between ancestral populations.  This might be
expected when modeling data with continuous ancestral population
structure. In contrast, the HapMap
data and, to a lesser extent, the ASW data exhibit more variability
across ancestral populations.  This may be a function of variable
heterozygosity within the distinct ancestral
populations~\cite{Conrad2006}. 
The higher observed discrepancy in ASW relative to the parallel recovered ancestral 
populations in the HapMap data may suggest that,
within this population of individuals with African and European ancestry, there is less
variability within the ancestral populations than in estimates of
these populations from non-admixed individuals (European and Yoruban
HapMap individuals, for example). This is interesting in light of
recent estimates of the effective population size of the ASW
population, which is an order of magnitude greater than either the
European or Yoruban effective population sizes~\cite{Excoffier2013}. 

A mismatch in observed and replication inter-individual similarity leading to a failure of this PPC may
be due to admixture linkage disequilibrium (LD).  Admixture LD arises because
ancestral recombination events will induce \emph{haplotype blocks}, or
stretches of the haploid chromosome, that are shared among many
individuals in a population~\cite{Daly2001,Gabriel2002}. Haplotype
blocks are population specific: the recombination hotspots, the
strength of correlation within the blocks, and the block length will
be specific to each population based on its recombination history.  A
few population specific features, including expansion rate and
migration rates, influence these haplotype block
characteristics~\cite{Greenwood2004}. As noted in earlier work
(e.g.,~\cite{Falush2003}), Haplotype blocks, and resulting admixture
LD, may account for greater than expected similarity among individuals
within these studies. Lower than expected similarity is likely due to
the proximity of the ancestral populations: if the ancestral
populations are difficult to distinguish with respect to allele
frequencies, then SNP-specific ancestral assignments will be
arbitrary, and within-population allele variance will be higher in the
observed data than is captured by the Bernoulli model and found in the
replicated data.

 \vspace{-3mm}
\subsection{Discrepancy in background LD}

Linkage disequilibrium (LD) is the non-random assortment of alleles
across the genome. LD occurs when alleles are not inherited
independently: the alleles at one genomic locus provide information
about the alleles at another locus. The process of recombination in
diploid chromosomes implies that alleles that are nearby on a
chromosome are inherited together unless a recombination event occurs,
creating dependencies in local genotypes
population-wide. Recombination is an infrequent event across a genome:
for a single offspring, there are on average a small number of
recombination events per chromosome, with high
variance~\cite{Fledel2011}. Recombination events
within a population lead to a block-like correlation structure of
genotypes across the genome, where polymorphisms adjacent in the
chromosome will be well correlated (referred to as \emph{background
  LD}, in contrast to possibly long-distance dependencies induced by
admixture LD). This correlation decays as the chromosomal distance
increases (although not uniformly). Although each study we analyzed
contains local correlation patterns, the admixture model assumes
independence of every SNP conditional on population ancestry.

We assessed the fitted admixture model for local correlation structure
among SNPs.  To do this, we built a discrepancy function (\emph{LD
  discrepancy}) that measures local dependencies among SNPs using
mutual information. Although we chose mutual information, the history
of statistics to quantify background LD is rich~\cite{Slatkin2008},
and there are other possibilities for this discrepancy statistic.
\emph{Mutual information} (MI) measures the dependence between two
random variables, here, the observed alleles at two loci, $x_{i,\ell,j}$
and $x_{i,\ell',j'}$. Specifically, MI quantifies the the reduction in
uncertainty about random variable $x_{i,\ell,j}$ given knowledge of the
state of $x_{i,\ell',j'}$ (or vice versa; MI is symmetric) for a pair of
discrete random variables~\cite{Cover1991} (Eqn.~1). If two SNPs are
independent (i.e., in \emph{linkage equilibrium}, or inherited
independently), MI will, theoretically, be zero. However, finite
samples imply that, even when two SNPs are in linkage equilibrium, the
MI may be greater than zero by chance.  Our discrepancy uses MI to
measure dependence between adjacent pairs of assayed SNPs.  We measure
the average MI between all pairs of genotypes assigned to the same
population. We checked lags between pairs of SNPs varying between $1$ and $30$, 
indicating $0$ up to $29$ intervening SNPs between the pairs of tested SNPs 
along the chromosome.


\begin{figure*}
\begin{center}
\includegraphics[width=7in]{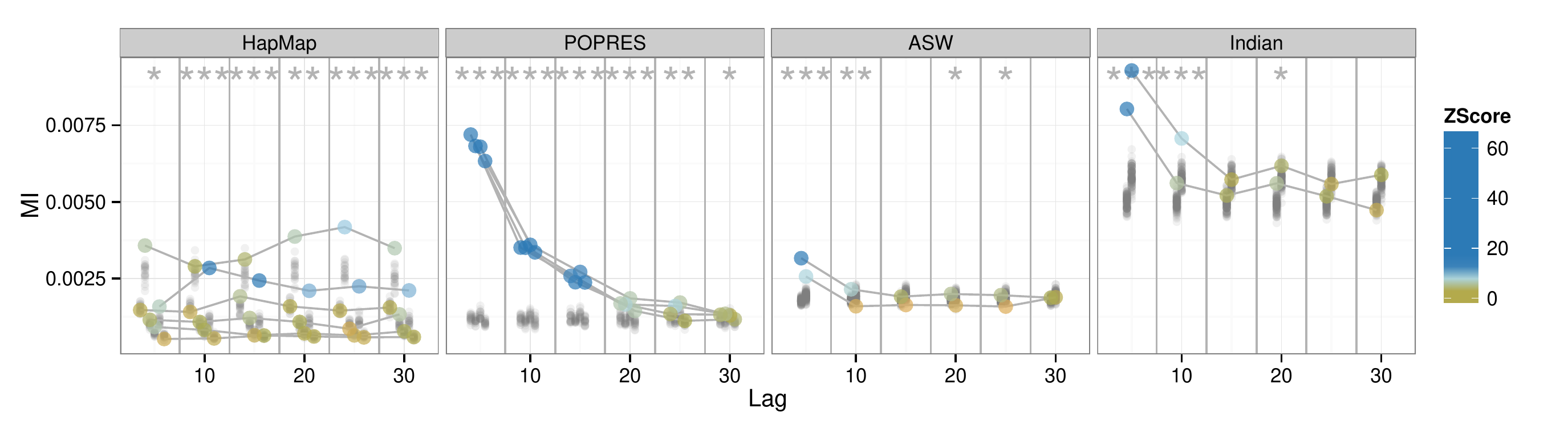}
\caption{{\bf LD discrepancy PPC across four studies.} The
  $x$-axis indicates the number of lag SNPs between pairs of SNPs over
  which mutual information is averaged. The $y$-axis represents the
  average mutual information between SNPs at varying lags. Observed
  values for each population are connected by lines across lags and
  colored by $z$-score. None of these studies are well fit across all
  SNP lags, highlighting the impact of the assumption of independence
  between SNPs.} 
\label{fig:mi-observed}
\end{center}
\vspace{-3mm}
\end{figure*}


When we applied this LD PPC to our data, we found that, 
across studies, the
models were generally misspecified at small lags (Figure~\ref{fig:mi-observed}). For all studies
but HapMap, the $z$-scores show that, as the lag grows, the MI values on the observed
data tend toward underestimates of MI in the posterior predictive distribution. The
$z$-scores for the POPRES data deviate substantially from a standard
normal for all lags indicating more observed background LD than
expected. The model's independence assumption across SNPs manifests in
replicated values that have near identical distributions across lags. Failure of this PPC may raise the possibility of using a subset of SNPs with low pairwise LD; however, admixture models rely on greater SNP densities to enable the separation of similar ancestral populations. Below we describe model extensions to address failure of the LD PPC.

We also looked at the observed discrepancies alone. In POPRES
and Indian data, we saw that background LD decays rapidly with
observed SNP lag on average~\cite{Altshuler2005}.  In
the POPRES data, we found the measurements of background LD to be
consistent across inferred populations (Figure~\ref{fig:mi-observed}). In the HapMap data,
background LD differs among distinct populations~\cite{Shifman2003};
this is expected because LD is impacted by population-specific
effects, including migration, deviations from random mating, and
selection~\cite{McEvoy2011}. In HapMap and, less so, the ASW data,
background LD is fairly uniform across different lags. It is possible
that this difference may be a function of the density of SNPs and SNP
ascertainment biases, which differs across studies and genotyping
methods, but we found that these results persist for lags up to 1000
adjacent SNPs. Note also that the HapMap data include almost 2.5 times
more SNPs than the POPRES data.

\vspace{-3mm}
\subsection{Discrepancy in reported ancestry}



In admixture models, it is often assumed that self-reported ancestries
provide no additional information above individual-specific ancestry
parameters.  We tested misspecification with respect to this
assumption. We developed a PPC for inferred population-specific values of
the fixation index $F_{ST}$ as compared to self-reported ancestry or
geography labels.  $F_{ST}$ measures the degree to which the variance
of allele frequencies within reported ancestries differs from the
variance of those alleles in the study conditioned on inferred
population assignment.  Using the fitted admixture model to partition
alleles into one of $K$ inferred ancestral populations, $F_{ST}$
measures whether or not subdividing these alleles by
individual-specific reported ancestry reduces the variance of those
allele frequencies.  The $F_{ST}$ is zero if no additional population
structure exists in the reported ancestries beyond what is already
recovered in the inferred ancestral populations.  Conversely, large
values of $F_{ST}$ indicate that the reported ancestries capture
additional structure---quantified by reduced allele variation within
individuals sharing reported ancestry---not found in the estimated
populations.  For a fitted admixture model with $K$ populations, we
evaluated this population-specific $F_{ST}$ discrepancy function by
computing $K$ $F_{ST}$ values, one for each inferred ancestral
population, with respect to reported ancestry (Eqn.~2).

\begin{figure}
\begin{center}
\includegraphics[scale=0.6]{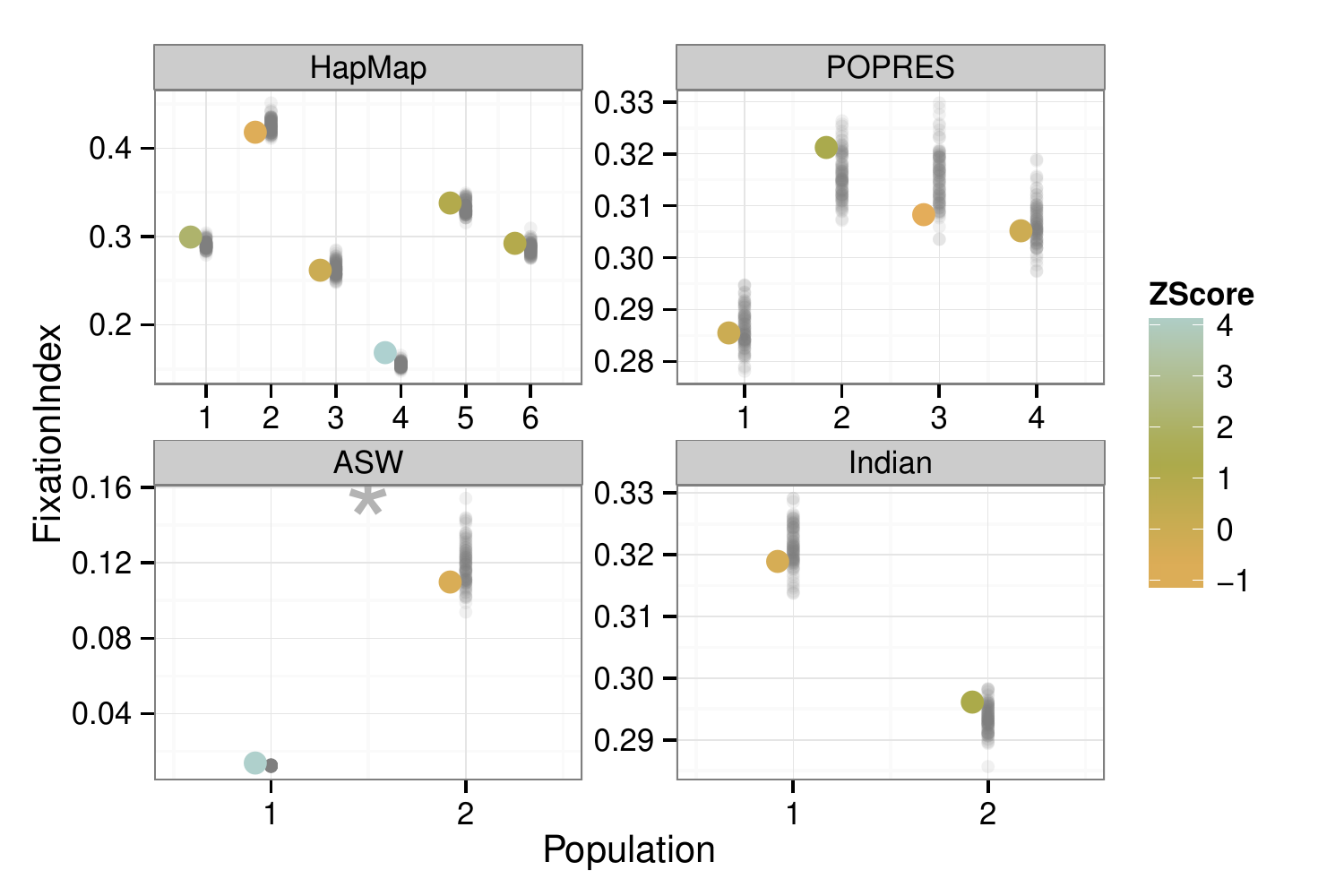}
\caption{ {\bf $F_{ST}$ discrepancy function across four studies.} The $x$-axis columns represent application to HapMap, POPRES, ASW, and Indian studies respectively. The $y$-axis represents the $F_{ST}$ value of a single ancestral population $k$ with respect to the self-reported ancestry information. Smaller values indicate a closer match between the reported and estimated ancestral populations. In ASW, it appears that number of ancestral populations $K$ is outside of the acceptable range.}
\label{fig:fst}
\end{center}
\vspace{-3mm}
\end{figure}

We computed the PPC for the $F_{ST}$ discrepancy, and we found that, except in ASW, the
estimated genetic ancestry often reflects all of the population
structure in the reported ancestry (Figure~\ref{fig:fst}). If we perform this PPC over many
different values of ancestral populations $K$, we found that this PPC
tends to fail for values of $K$ that are inadequate to explain
variation in the observed data (Figure~S3). This discrepancy appears
to be, in these four studies, an effective approach to evaluate the
range of well-specified numbers of ancestral populations in the
admixture model. Based on previous observations, we hypothesize that LD 
is inducing a phantom population in the ASW study that is not captured using two ancestral populations~\cite{WTCCC}. This would lead to the failure of this PPC, although the reference individuals included are thought to capture the two admixed ancestral populations of the ASW individuals\cite{1000genomes}.

Note that the observed $F_{ST}$ varied considerably across studies
(Figure~\ref{fig:fst}).  As with other discrepancy functions, HapMap
and ASW show greater variability of observed $F_{ST}$ values across
populations than POPRES and Indian due to greater heterogeneity of
ancestral populations.  The two ASW populations have consistently
lower observed $F_{ST}$ values, possibly suggesting that the fitted
admixture model captures most of the information in the reported
ancestries (ASW, CEU, and YRI).  Indeed, $F_{ST}$ is undefined for
several populations within the ASW models because all the alleles
assigned to those populations were from individuals with a single
reported ancestry.  In the context of the replicated data, however, this conclusion is shown to be incorrect: there is latent structure in the data that is not captured well by two ancestral populations. The Indian data, with $14$ distinct reported
ancestries, have higher overall $F_{ST}$ values: individuals within
each reported ancestry have a range of admixture between Ancestral
North Indian and Ancestral South Indian~\cite{Reich2009}, and reported
geographic labels are, unsurprisingly, a poor indicator of admixture proportions across
these individuals.

\vspace{-5mm}
\subsection{Discrepancy in uncertainty in ancestral population assignments}

We chose these four studies because they represent distinct patterns of
admixture (Figure~\ref{fig:rainbow-diagram}): in the HapMap data, most
individuals have a majority of genomic ancestry explained by one or
two populations; in POPRES, although regional variation in
ancestry proportions is evident, most individuals have substantial
ancestry contributions from all four populations.  We measured the
degree of uncertainty of individual-specific population assignments to
determine whether this uncertainty is characteristic of the model
parameters or evidence of model misspecification.

The discrepancy function for this task quantifies the average entropy,
or uncertainty, of the ancestral population assignment for alleles in
the fitted model.  We used the fitted admixture model to compute
estimates of the conditional probability of the assignment of each SNP
allele to each ancestral population $k \in \{1,\dots,K\}$. We then
computed, for each population, the average entropy of this conditional
probability across individuals and SNPs, including SNPs assigned to
the same population $k$ (Eqn.~3). 

\begin{figure}
\begin{center}
\includegraphics[scale=0.6]{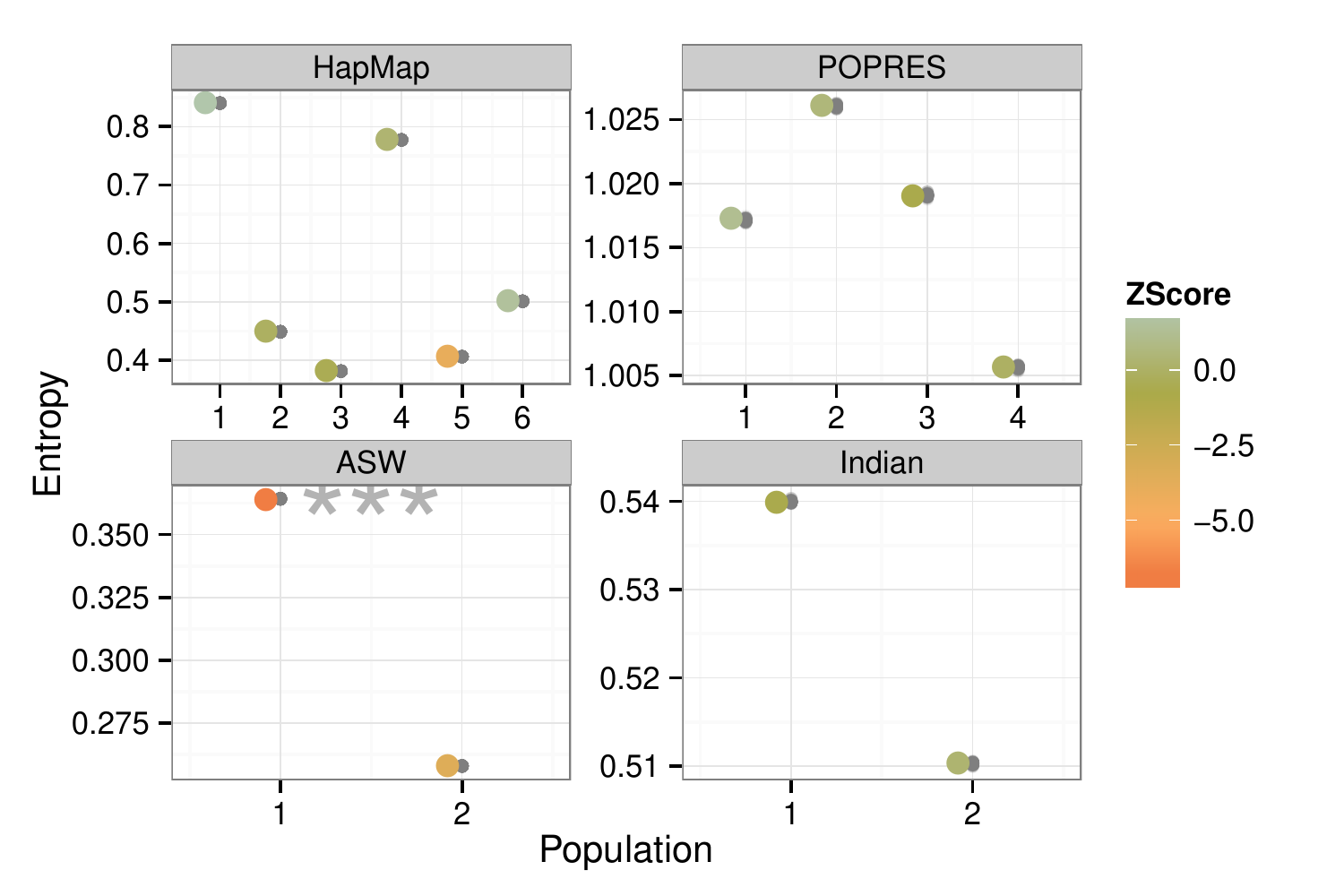}
\caption{ {\bf Average entropy discrepancy function across four studies.} The $x$-axis represents the number of ancestral populations in the fitted admixture model; panels represent application to HapMap, POPRES, ASW, and Indian studies. The $y$-axis represents the average entropy of the posterior distribution over populations for each allele. Larger absolute value entropy represents greater uncertainty over population assignments. Replicated values are shown as smaller gray circles. This discrepancy function shows small variance among replicates and finds a misspecified model for the ASW study.}
\label{fig:entropy}
\end{center}
\vspace{-3mm}
\end{figure}

We performed PPCs with this entropy discrepancy, and found that this model is grossly misspecified for the ASW study with respect
to uncertainty in allele-specific ancestry assignments
(Figure~\ref{fig:entropy}); the remaining three studies did not indicate problems with this PPC. This result is interesting considering the low
variation in average entropy among the replicated data; indeed, the
replicated points appear near-identical at the scale of the
visualization. As with the $F_{ST}$ PPC, we hypothesize that the failure of this PPC on the ASW population may be due to the additional ancestral structure induced by LD that is not captured well with two ancestral populations.

The success of this PPC across three studies is also unexpected 
when considering the observed discrepancies alone.
 We found that the POPRES and Indian studies showed high
observed average entropy relative to the other studies along with low variation
in observed average entropy across populations, indicating a high degree of
uncertainty in population assignments (Figure~\ref{fig:entropy}).
This uncertainty does not appear to be caused by lack of convergence
in the parameter estimates: this same behavior is observed for models
fit with ten times as many EM iterations. In contrast, the HapMap and
ASW studies, which have distinct ancestral populations, have greater
variance of average entropy across populations within $K$, indicating
greater uncertainty in some population assignments relative to others.

\vspace{-3mm}
\subsection{Discrepancy in correcting for population structure in genome-wide association studies}

A final discrepancy function we considered is the value of using
estimates of individual specific admixture proportions to correct for
population structure in association
studies~\cite{Pritchard2000b,Price2006,Hoggart2003,Satten2001,Devlin1999}. \emph{Association
  mapping} uncovers associations between genetic variants (e.g., SNPs)
and traits (e.g., height, disease status, cholesterol levels). It is
well-appreciated that correcting for population structure in
association mapping is essential because latent structure leads to
false positive associations when alleles that have different
frequencies across populations are mapped to traits with differential
rates across populations~\cite{Patterson2004,Marchini2004}. We note
that, in the original manuscript describing the use of admixture model
parameters to correct for population stratification in association
studies~\cite{Pritchard2000b}, they effectively performed the same PPC
as we applied here without describing it as a posterior predictive
check.

For this task, we evaluated the effect of the fitted admixture
parameters on the distribution of $\log_{10}$ Bayes factors (BFs)
comparing the null hypothesis of no association between a SNP and the
trait controlling for population assignment, and the alternative
hypothesis of an association between a SNP and the trait, controlling
for population assignment (following prior work~\cite{Pritchard2000b};
see Methods). We randomly generated binary traits for each study using
a population $k$ in the generative model, creating phenotypes with
different rates within our estimated populations but with no explicit
association with any SNP. Thus, controlling for population structure,
any significant association will be a false positive. The value of the
discrepancy for a specific population is computed as the maximum
$\log_{10}$BF over all SNPs. We then computed the $z$-score of the
$\log_{10}$BFs of the observed data associations with respect to the
$\log_{10}$BFs from the replicated data.

\begin{figure}
\begin{center}
\includegraphics[scale=0.6]{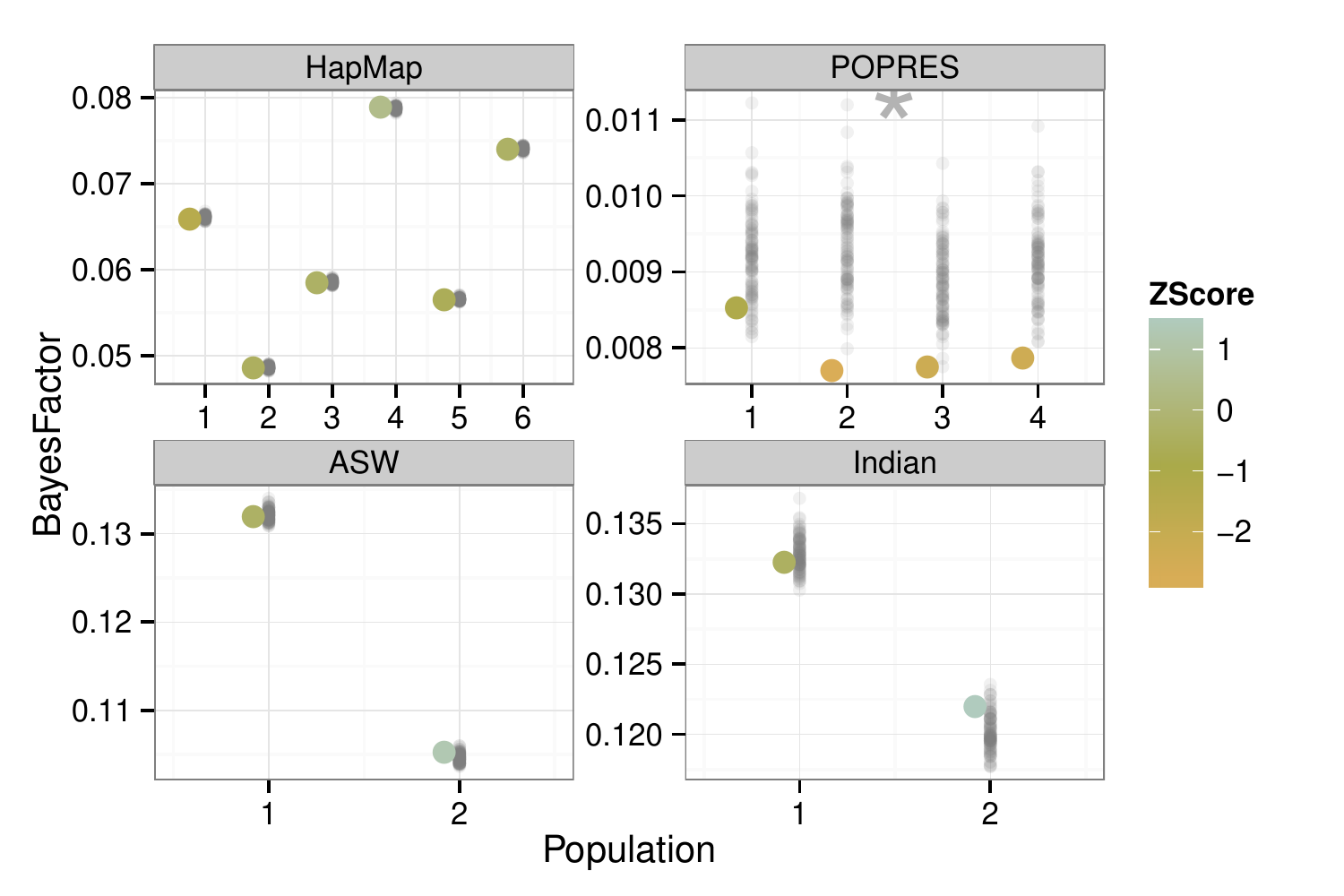}
\caption{ {\bf Association mapping correction PPC across four studies.} The $x$-axis represents application to HapMap, POPRES, ASW, and Indian studies respectively. The $y$-axis represents maximum $\log_{10}$BFs. Maximum $\log_{10}$BFs from replicated genomes are shown with smaller gray points. Colors of the observed data discrepancy values represent the $z$-score with respect to the replicated samples from fitted admixture models. Stars indicate deviation from normality in $z$-scores. The POPRES data appears to deviate from the estimated posterior predictive distribution for association mapping.}
\label{fig:gwas}
\end{center}
\vspace{-3mm}
\end{figure}

We performed this mapping PPC and found that the observed maximum
$\log_{10}$BFs are generally within the expected range when sampling
alleles from the fitted model, which provides additional confidence in
our ability to reject false positives (Figure~\ref{fig:gwas}). The variation in the POPRES
replicates highlights why the PPC fails: controlling for fine levels
of population structure with noisy discrete estimates is not effective
control for structure. It is counterintuitive, though, that the
four populations in POPRES have lower than expected observed
correction $\log_{10}$BFs.

The observed discrepancies alone were also informative We found that the maximum
$\log_{10}$BF across studies was small ($< 0.14$) indicating that this
approach to correcting for population structure is broadly effective
at avoiding false positive associations.  As in previous
discrepancies, the variability in maximum $\log_{10}$BF between
populations was greater for HapMap and ASW data than POPRES and Indian
data, reflecting greater distinction between populations in the first
two studies. The maximum $\log_{10}$BFs for POPRES are smaller than
the other studies, reflecting less significance in tests for
association between SNPs and populations; this is not surprising given
the homogeneity of the inferred populations and the corresponding
randomly generated phenotypes.

\vspace{-3mm}
\subsection{Summarizing PPC results within study}

We turn our attention to summarizing the results from the five PPCs
for each of our genomic studies.  Our results across PPCs tell a
complex story for each study that indicates specific misspecified
assumptions.

\vspace{-4mm}
\paragraph{HapMap phase 3.}

Across our application of PPCs to the HapMap phase 3 data, we found
that the background LD discrepancy PPC indicated a gross model
misspecification, highlighting the negative impact of the assumption
of independent SNPs. Other PPCs did not find misspecification with
$K=6$ on these data. Two ways to address this model misspecification
with respect to background LD in these well-separated ancestral
populations would be to a) prune the SNP data drastically to select a
near-independent set of SNPs from which ancestry may be estimated, or
b) model background LD explicitly (see Discussion).

\vspace{-4mm}
\paragraph{European samples.}

Across our application of PPCs to the POPRES data, we found that the
admixture model with $K=4$ ancestral populations generally indicated
similar variation in discrepancy within and across populations.  Many
of the failures of the PPCs on these data could be used together to
highlight model misspecificiation, as the continuous structure of the
ancestral populations is not captured well by the discrete populations
assumed by an admixture model. A possible solution is to model latent
structure for these data with a continuous population model (e.g.,
principal components-based~\cite{Patterson2006,Price2010}).

\vspace{-4mm}
\paragraph{African Americans.}

PPCs on the ASW data with $K=2$ showed
that the ancestral population corresponding to European ancestry was
well captured in the observed data with respect to the replicated
data, but the population corresponding to Yoruban ancestry was often
misspecified for the observed data with respect to the replicated
data; the $F_{ST}$ PPC and the entropy PPC show this differential fit.  For these data,
explicitly modeling admixture LD may enable a better fit with two
ancestral populations, eliminating the need to use an additional
ancestral population to control for the effects of long distance
haplotype block correlations~\cite{WTCCC}.

\vspace{-4mm}
\paragraph{Continental Indians.}

With PPCs on the Indian data with $K=2$, we found that the
failure of the average entropy PPC indicates that the underlying
estimates of the two ancestral populations have substantial uncertainty.
Relying on these estimates to characterize ancestral population
allele frequencies or determine admixture proportions for each
individual is unjustifiable. These data may also benefit from using a
continuous model of ancestral populations because of the difficulty of
differentiating these two fairly proximal ancestral populations many
generations after the admixture events~\cite{Reich2009}.


\section{Discussion}

We have developed posterior predictive checks (PPCs) for analyzing
genomic data sets with the admixture model.  We have demonstrated that
the PPC---estimating the posterior predictive distribution and
checking the likelihood of the true observed data under this distribution---gives a valuable
perspective on genetic data beyond statistical inference of model
parameters.  In the research literature, fitted admixture models are
often accompanied by a `just so' story to explain the inferred
parameters and how they are reflective of ancestral
truth~\cite{Rosenberg2002}.  The model may suggest these hypotheses,
but only conditioned on the model being a good fit for the observed
data.  PPCs check this assumption of good fit, giving weight to the hypotheses by
confirming that the underlying assumptions do not oversimplify the 
existing structure in the observed data.  In this paper, we developed PPCs for the admixture model,
designing biological discrepancy functions to quantify the effect of
the model assumptions on interpreting and using the estimated
parameters for downstream analyses.

Statistical modeling of genetic data requires us to balance the
complexity of the model with its capacity to capture the data at
hand. As examples of limitations, we may not have enough data to
support an overly complex model, or the model class that that we want
to fit may be too complex given our computational constraints. Thus, we
support the iterative practice of fitting the simplest model (i.e.,
the one we fit here), checking whether a higher resolution model is
needed, and then improving the model only in the ways that result in
more reliable interpretations of the results.  PPCs can drive this
process of targeted model development, pointing us towards enriched
Bayesian admixture models along gradients that quantifiably improve
their performance for the exploratory tasks that matter.  With this
practice in mind, we revisit the PPCs described above and discuss how
we might enrich the simple admixture model to address its misspecified
assumptions.

Many population studies have applied admixture models to explore and
quantify genetic variation between individuals within and across
ancestral populations~\cite{Rosenberg2002,Jorde2004,Serre2004}; these
analyses may benefit from the inter-individual PPC.  For studies
where this PPC indicates misfit, prior work has adapted the admixture
model to control admixture LD by explicitly modeling haplotype blocks
for each ancestral population instead of modeling each SNP
separately~\cite{Falush2003}. In particular, the SNP-specific ancestry
assignment $z$ variables for each individual are modeled by a Markov
chain, where the probability of transitioning to a different ancestral
population from one position to the next has an exponential
distribution. This specifies a Poisson process describing the length
of haplotype blocks across the chromosome, with global rate parameter
$r$.

Many studies have noted that background LD may lead to phantom
ancestral populations~\cite{WTCCC}; applying admixture models to
genomic data that contain background LD may find the SNP
autocorrelation PPC useful. After identifying model misspecification
using our background LD discrepancy function, we could extend the
admixture model to explicitly capture background LD. Above we
described a Markov model on the $z$ variables. It assumes that,
conditional on ancestral population assignment, genotypes are
independent.  Extending this idea, SABER~\cite{Tang2006} implements a
Markov hidden Markov (MHMM) model to capture both haplotype blocks and
background LD by adding a Markov chain across the population-specific
allele frequencies in $\beta$.  Others have further extended this
model in various ways, including estimating recombination events
explicitly in the MHMM~\cite{Sankararaman2008}.


The $F_{ST}$ discrepancy function effectively checks for a
misspecified number of ancestral populations $K$. The ubiquitous
problem of selecting a number of ancestral populations is, arguably,
the most substantial hurdle to overcome in applying admixture models,
or latent factor models generally, to
data~\cite{pritchard2000inference,Rosenberg2002,Price2006}. Methods
and statistics have been proposed to evaluate the proper number of
latent ancestral populations, often motivated by
$F_{ST}$~\cite{pritchard2000inference,Evanno2005}; additionally,
nonparametric Bayesian models estimate the posterior probability for
each $K$~\cite{Huelsenbeck2007,Teh2006}. We propose a PPC with the
$F_{ST}$ discrepancy for general use in evaluating appropriate ranges
of the number of ancestral populations for a specific study. A simple
adaptation of the model to correct for a failure of this PPC is to
change the number of ancestral populations $K$ (Figure~S3).

There are also explicit model adaptations that will affect the
$F_{ST}$ of the inferred ancestral populations.  For example, one can
build hierarchical models that allow the sharing of allele frequencies
across populations for some SNPs; this was implemented in the
\emph{structure} 2.0 model, which includes a hierarchical component to
allow similar allele frequencies across ancestral populations (the
so-called $F$ model)~\cite{Falush2003}. A second example is from the
topic model literature (similar models applied to modeling text
documents), where the ancestral populations are captured in a
tree-structured hierarchy~\cite{Blei2003b, Adams2010}. In the
corresponding admixture models, the root node would include SNPs that
have shared allele frequencies across all ancestral populations; at
the leaves, the population-specific allele frequencies would include
SNPs that have a frequency in that population that is different than
the frequency in all other ancestral populations (referred to as
\emph{ancestry informative markers}~\cite{Rosenberg2003}).  



Previous population studies have explored and interpreted the
population-specific SNP frequencies estimated by admixture
models~\cite{Ardlie2002,Pickrell2012,Gautier2013}; almost all
applications of this admixture model have used MAP estimates of
ancestry assignments to determine the proportion of admixture in
individuals~\cite{Rosenberg2003,Engelhardt2010}.  The average entropy
PPC will check model misspecification for ancestry assignment, and has
implications for interpreting estimates of SNP frequencies.  To adapt
the model to this misspecification, the hyperparameters for the
Dirichlet-distributed allele-specific ancestry assignments may be
changed.  (We and others set to $\alpha =
1$~\cite{pritchard2000inference}, giving equal weight to all possible
contribution across ancestries for each SNP.) In particular, we might
give higher weight to admixture proportions near $0$ and $1$ by
setting $\alpha < 1$ for studies where we expect low levels of
admixture (e.g., the HapMap data). The equivalent change for the
hyperparameters in the population-specific allele frequency parameters
would encourage for allele frequency spectrums that more closely match
what we find in natural populations~\cite{Nelson2012}. Another
relevant model adaptation would be to modify the distribution of a SNP
to be not Bernoulli but instead Poisson~\cite{Lee1999},
normal~\cite{Rasmussen2000}, or something more
sophisticated~\cite{Yang2012,Paisley2012}. We emphasize that, though
these extensions seem reasonable, the PPC with this discrepancy found
little need to modify the admixture model assumptions in our current
studies. The exception to this point is the ASW study, although we 
hypothesize that correcting for background LD as suggested above will 
address this misspecification. 

We believe that all model-based methods to control for population
stratification in association mapping will benefit from application of the
mapping PPC, including linear mixed models and non-generative methods
such as EIGENSTRAT~\cite{Price2006,Kang2010}.  Failure of the
association mapping PPC indicates that the estimates of population
structure are insufficient to correct for the confounding latent
structure in the individuals.  There are many directions to consider
for mitigating this type of model misspecification.  As examples, one
may use larger numbers of estimated principal components or ancestral populations, use
alternative approaches to specifying the latent structure variables,
or correct for structure that are estimated on local regions of the
genome.  This same discrepancy function---replacing $z$ with the
estimated random effect from linear mixed models---would be useful in
quantifying model misspecification for these alternative methods for
association mapping in the presence of confounding population
structure~\cite{Listgarten2012,Zhou2012,Runcie2013}.


Applied statisticians develop models to capture the biological
complexity of their data.  To form hypotheses from these models,
however, we need assurances that the data can support them.  PPCs
provide a simple mechanism to quantify when a model is sufficient or when it needs additional structure to support downstream analysis.  While we have focused
on the admixture model, the PPC methodology applies to any probabilistic
model of data.  For example, we believe there could be a substantial
role for PPCs in evaluating demographic models. As we continue to
collect complex genomic data, we continue to develop complex models to
explain them.  Equally important to building our repertoire of
statistical models for analyzing genomic data is to build our
repertoire of ways to check those analyses.



\vspace{-5mm}
\subsection{Genomic study data}
We downloaded the HapMap phase 3 release \#3 genotype data from the NCBI website~\cite{Altshuler2010}. We downloaded the POPRES data from dbGaP (accession number phs000145) and filtered these genotype data as described in prior work~\cite{Novembre2008,Engelhardt2010}. We downloaded the ASW genotype data from 61 individuals, $32$ CEU individuals, and $37$ YRI individuals from the 1000 Genomes Project Phase 1~\cite{1000genomes} from the 1000 Genomes FTP server~\footnote{\tt ftp://ftp.1000genomes.ebi.ac.uk/vol1/ftp/phase1/analysis\_results} and pruned these data by randomly sampling 1\% of SNPs with MAF $\geq 5\%$. We downloaded the Indian data from the Reich Lab FTP server~\cite{Reich2009} and processed it as in previous work~\cite{Engelhardt2010}.

\vspace{-3mm}
\subsection{Admixture model and parameter estimation}
We used a standard admixture model in this work, as implemented in the \emph{Structure} software~\cite{pritchard2000inference}, although we built our own software to fit this model to large scale data as described below. We represented ancestral population $k \in \{1,\dots,K\}$ with a Dirichlet-distributed random variable $\phi_{\ell, k}$, for each single nucleotide polymorphism (SNP) $\ell \in \{1,\dots,L\}$, over all possible alleles; we assumed exactly two alleles for each SNP, so this simplified to a beta distribution.
We represented individual-specific ancestry proportions for individual $i \in \{1,\dots,n\}$ as a Dirichlet-distributed variable $\theta_i$ over $K$ populations. To generate a diploid genotype $i$ from this model, for each SNP $\ell$ we sampled two multinomial variables with parameter $\theta_i$, $z_{i,\ell, 1}$ and $z_{i,\ell,2}$, one for each allele copy.
We then sampled alleles $x_{i, \ell, 1}$ and $x_{i,\ell, 2}$ from the distributions $\phi_{\ell, z_{i,\ell, 1}}$ and $\phi_{\ell, z_{i, \ell, 2}}$, respectively. The generative model has the following form:
\vspace{-1.5mm}
\begin{eqnarray*}
\theta_i &\sim& Dir_K(\alpha)\\
\phi_{\ell, z_{i,\ell,j}} &\sim& Beta(\gamma,\gamma)\\
z_{i,\ell, j} &\sim& Mult(\theta_i) \mbox{ for }j \in \{1,2\}\\
x_{i,\ell,j} &\sim& Mult(\phi_{\ell, z_{i,\ell,j}}) \mbox{ for }j \in \{1,2\},
\end{eqnarray*}
where $j \in \{1,2\}$ represents the two allele copies.  For simplicity, we set $\alpha = 1$ and $\gamma = 1$, indicating a uniform prior on the ancestry proportions for each individual and the site-specific allele frequencies for each population~\cite{alexander2009fast}.
The likelihood for this model has the form:
\begin{align*}
P(\bx, \bz | \bm{\theta}, \bm{\phi}) = & \prod_i^n \prod_\ell^L \prod_{j \in \{1,2\}}
\phi_{\ell, z_{i, \ell, j}}^{x_{i, \ell, j}} (1-\phi_{\ell, z_{i, \ell, j}})^{1-x_{i, \ell, j}} \theta_{i,  z_{i, \ell, j}}.
\end{align*}
Without loss of generality, we represent a heterozygous SNP (encoded as a 1) in the data as $x_{i,\ell,1} = 1$ and $x_{i,\ell,2} = 0$, where $1$ and $0$ are the two alleles at that site~\cite{pritchard2000inference}.

We estimated these parameters and latent variables with an expectation maximization (EM) algorithm that alternates between i) (E-step) estimating the posterior mode for population assignments of individual alleles $Z$ given estimates for $\bm{\phi}$ and $\bm{\theta}$, and ii) (M-step) maximizing the individual- and population-level parameters given those posterior modes of $Z$~\cite{Dempster1977,alexander2009fast}.
The update equations for the E-step, estimating the posterior mode for population assignment for an allele, are
\begin{align*}
q(z_{i,\ell, j}=k | x_{i,\ell,j} = 1, \phi_{\ell,k}, \theta_{i,k}) & \propto \theta_{i,k} \phi_{\ell, k} \\
q(z_{i,\ell, j}=k | x_{i,\ell,j} = 0, \phi_{\ell,k}, \theta_{i,k}) & \propto \theta_{i,k} (1 - \phi_{\ell, k}).
\end{align*}
The update equations for the M-step, estimating the model parameters for $n$ individuals and $L$ SNPs given the posterior mode for $\Z$, are:
\begin{eqnarray*}
\phi_{\ell,k} & =& \frac{1}{2n} \sum_{i=1}^n \sum_{j \in \{1,2\}}
q(z_{i,\ell,j}=k)^{\1[x_{i ,\ell,j} = 1]} \\
& & (1 - q(z_{i, \ell, j}=k))^{\1[x_{i, \ell, j} = 0]} \\ 
\theta_{i,k} & = & \frac{1}{2L} \sum_{\ell=1}^L \sum_{j \in \{1,2\}} q(z_{i,\ell,j}=k). \\
\end{eqnarray*} 
We initialized the population-specific minor allele frequency parameters $\phi_{\ell, k}$ to the empirical proportion of the minor allele in the training data, plus a uniform random variable $u \sim \mathcal{U}(0, 0.1)$, truncating extreme values so that $0.05 < \phi_{\ell, k} < 0.95$.
To initialize population proportions for each individual, we drew $K$ uniform random variables $u_k \sim \mathcal{U}(0, 1)$ and set $\theta_{i,k} = u_k/ \sum_{k'}^K u_{k'}$. We iterated between these E- and M-steps for 1000 iterations. We also ran a number of models to 10,000 iterations, but found no substantial differences in the fitted models, supporting convergence of the parameter estimates in 1000 iterations.

\vspace{-3mm}
\subsection{Discrepancy: Inter-individual similarity for IBD}
We measured similarity between individuals (identity by descent) with respect to population $k$ by counting the number of alleles that are shared between individuals with population assignment $z_{i,\ell,j} = k$ (i.e., the Manhattan distance) and dividing by the total number of alleles with population assignment $z_{i,\ell,j} = k$~\cite{Weir1984}.
The value of the discrepancy function is the average proportion of shared alleles over all pairs of individuals. We ignored pairs of genomes that shared fewer than $500$ alleles assigned to a population $k$.

\vspace{-3mm}
\subsection{Discrepancy: Mutual information for background LD}
We calculated population-specific MI for each pair of adjacent SNPs within a lag of all integers $m \in \{2,\dots,30\}$.  This window is with respect to the ordering of the SNPs in our filtered data set based on chromosomal position, although the actual distance in base pairs between SNPs varies dramatically across studies and SNP pairs. As with the inter-individual similarity, we only compared pairs of SNPs if they shared population assignments.

This discrepancy function calculates mutual information (MI) quantified in \emph{bits} between the observed alleles at two adjacent loci separated by $m-1$ SNPs, $x_{i,\ell,j}$ and $x_{i,\ell',j'}$. For haploid genotypes, both generated from the same ancestral population $k$, we computed:
\begin{eqnarray}
\nonumber I(x_{i,\ell,j};x_{i,\ell',j'}) &=& \sum_{x_{i,\ell,j} \in \{0,1\}} \sum_{x_{i,\ell',j'} \in \{0,1\}} p(x_{i,\ell,j},x_{i,\ell',j'}) \\
&&\log_2 \left(\frac {p(x_{i,\ell,j},x_{i,\ell',j'})} {p(x_{i,\ell,j})p(x_{i,\ell',j'})}\right).
\end{eqnarray}
As calculating this statistic is computationally intensive, we computed MI for this $30$ SNP window only over the first 10,000 SNPs in each study.

\vspace{-3mm}
\subsection{Discrepancy: Distance between estimated ancestral populations and geographic labels}
We computed the $F_{ST}$ statistic on alleles assigned to one ancestral population relative to the geographic labels using Wright's estimator of $F_{ST}$, which considers single alleles at each genomic locus~\cite{wright1969evolution}.
In our data, each individual has exactly one geographic label $g$, but the individual's alleles are assigned to possibly many ancestral populations.
The probability of an allele for SNP $\ell$ and population $k$ is computed as $\bar{p}_{k, \ell} = \frac{1}{N_{k, \ell}}\sum_i \sum_j^2 x_{i, \ell,j} I_{z_{i, \ell,j} = k}$, where $N_{k, \ell}$ is the total number of alleles assigned to population $k$ and $I_{\cdot}$ is an indicator function.
We further partition these population-specific probabilities by geographic label $g$: $\bar{p}_{k, g, \ell} = \frac{1}{N_{k, g, \ell}}\sum_{i \in g} \sum_j^2 x_{i, \ell,j} I_{z_{i,\ell,j} = k}$, where $N_{k, g, \ell}$ is the total number of alleles assigned to population $k$ for individuals with geographic label $g$, which may be zero, in which case this probability is set to zero.
With these probabilities in hand, and assuming Bernoulli distribution of alleles (so that the variance of these distributions is estimated from the expectations), we calculate fixation index $F_{ST}$ as
\begin{align}
F_{ST} = & \frac{ \frac{1}{R_{k}} \sum_{g=1}^{|g|} (\bar{p}_{k, \ell} - \bar{p}_{k, g, \ell})^2 }{ \bar{p}_{k, \ell} (1-\bar{p}_{k, \ell})},
\end{align}
where $R_{k}$ is the total number of geographic regions with non-zero allele counts for population $k$.

\vspace{-3mm}
\subsection{Discrepancy: Average entropy}
We computed the average entropy discrepancy function as the average entropy for each estimated ancestral population over all alleles assigned to a population. Given estimates of the distribution of ancestral populations for an individual $i$, $\hat{\theta}_{i,\cdot}$, and the probability of the minor allele for a SNP across populations, $\phi_{\ell, \cdot}$, we calculated the posterior probability of each population $k$ for the observed alleles $x_{i, \ell,\cdot}$ in that individual's genome.
The entropy is 
\begin{align}
\mathcal{H}(z_{i,\ell,\cdot} | x_{i,\ell,\cdot}) = -\sum_{k=1}^K p(z_{i,\ell,\cdot}=k | x_{i,\ell,\cdot}) \log_2 p(z=k | x_{i, \ell,\cdot}). 
\end{align}
Then we computed the discrepancy as the average entropy for each ancestral population $k$ over all alleles assigned to a population. 

\vspace{-1mm}
\subsection{Discrepancy: Correcting latent structure in association mapping}
For each model and each ancestral population $k$, we generated a binary phenotype vector of length $n$, simulating a condition that is associated with that population.
In particular, for the $k^{th}$ phenotype vector, we sampled a Bernoulli random variable $c_i$ for each individual with probability $0.5 \theta_{k,i} + 0.1 (1 - \theta{k,i})$, so individuals with ancestry in population $k$ are more likely to exhibit the phenotype.
To compute the discrepancy function for the GWAS association controlling for ancestry,
we computed, for each SNP, the $2\ln$ Bayes factor (BF) for each SNP as the ratio of the likelihood of the genotype given the phenotype and the population assignment of the SNP ($p(x_i | c_i, z_i)$), and the likelihood of the genotype given the population assignment ($p(x_i | z_i)$)~\cite{Pritchard2000b,Stephens2009}.
We used a beta-binomial model with symmetric smoothing parameter $0.1$ for the generalized linear model of association.
The value of the discrepancy for a population $k$ is the maximum $2\ln$ BF over all SNPs.
For each population $k$, we sampled random phenotypes ten times using that population as the risk factor and averaged the result of the discrepancy function over these ten samples.

\vspace{-3mm}
\subsubsection{Diploid data}
\vspace{-3mm}
Diploid data complicates some discrepancy functions, because, conditionally, each of the two copies of an allele may be generated from (or, equivalently, assigned to) different ancestral populations.
As in the original specification of the admixture model, we divided single ternary observations $x \in \{0, 1, 2\}$ into the sum of two binary observations $x_1, x_2 \in \{0,1\}$. Each binary observation has a separate latent population assignment $z_1, z_2$. Two of our discrepancy functions (i.e., inter-individual similarity, MI) compared pairs of SNPs and were modified to account for as many as two population assignments per SNP.

Let $z_{a1}, z_{a2}, z_{b1}, z_{b2}$ be the first and second hidden assignment variables at SNPs $a$ and $b$, and $x_{a1}, x_{a2}, x_{b1}, x_{b2}$ be the observed alleles associated with those hidden variables.
There are four possible comparison scenarios, depending on how many of the latent population assignment variables are set to $k$ among these two SNPs.
\vspace{-3mm}
\begin{enumerate}
\item {\bf No Match.} If at least one of the positions has no hidden variables equal to $k$, there are no relevant comparisons.
\item {\bf Single Match.} If exactly one hidden variable at both positions is equal to $k$, the two observations associated with those matching variables are comparable. 
For example, if $z_{a1} = k$ and $z_{b2} = k$, but $z_{a2}$ and $z_{b1}$ are set to two other populations, then we compare $x_{a1}$ with $x_{b2}$.
\item {\bf Single-Double Match.} If one location has one latent assignment variable equal to $k$, and the other location has both latent assignment variables equal to $k$, there are two possible comparisons.
For example, if $z_{a1} = k, z_{a2} = k$ and $z_{b2} = k$, we compare $x_{a1}$ with $x_{b2}$ and $x_{a2}$ with $x_{b2}$.
\item {\bf Double-Double Match.} In the case where all four hidden variables are equal to $k$, we have four possible comparisons: $x_{a1}$ with $x_{b1}$, $x_{a1}$ with $x_{b2}$, $x_{a2}$ with $x_{b1}$, and $x_{a2}$ with $x_{b2}$.
\end{enumerate}

\vspace{-3mm}
\subsection{Posterior predictive checks}
We generated replications of the observed data by sampling from the fitted model distributions:
\begin{align}
x_{i, \ell, j}^{REP} & \sim \mathcal{M}ult(\hat{\phi}_{\ell, \hat{z}_{i, \ell, j}}).
\end{align}
For a discrepancy function of alleles and their population assignment variables, $f(\bx, \bz)$, we calculated $f(\bx^1, \bz), \dots, f(\bx^R, \bz)$, holding the latent variables fixed. For each of our PPCs, we set $R = 100$, except for the intra-individual similarity that considers all pairs of replicates, where we used $R=30$.
We calculated the mean $\mu_k$ and standard deviation $\sigma_k$ of the values of the discrepancy function over all $R$ replications for each of $K$ ancestral population.
We then computed an empirical $z$-score to compare the function values for the observed data $\bx$ to the mean and standard deviation of the function values for the replications: $\sigma_k^{-1} (f(\bx, \bz) - \mu_k)$. 
Given these population-specific $z$-scores, we estimated the significance of their deviation from a standard normal distribution by computing a Bayes factor (BF) as the ratio of density of the $z$-scores under a the normal distribution with MLE parameters over the likelihood of the $z$-scores under the standard normal. We report this BF in all figures~--~three stars for $2\log BF > 10$, two stars for $2\log BF > 6$, and one star for $2\log BF > 2$~--~where a larger BF indicates greater distance from the standard normal for these observed $z$-scores.

\section*{Acknowledgements}
The authors would like to acknowledge the groups that made these data publicly available: HapMap Consortium, Glaxo-Smith Klein, 1000 Genomes Project, and David Reich. BEE was funded through NIH R00 HG006265.

\bibliographystyle{plos2009}
\bibliography{bio_ppcs}

\begin{thebibliography}{10}
\providecommand{\url}[1]{\texttt{#1}}
\providecommand{\urlprefix}{URL }
\expandafter\ifx\csname urlstyle\endcsname\relax
  \providecommand{\doi}[1]{doi:\discretionary{}{}{}#1}\else
  \providecommand{\doi}{doi:\discretionary{}{}{}\begingroup
  \urlstyle{rm}\Url}\fi
\providecommand{\bibAnnoteFile}[1]{%
  \IfFileExists{#1}{\begin{quotation}\noindent\textsc{Key:} #1\\
  \textsc{Annotation:}\ \input{#1}\end{quotation}}{}}
\providecommand{\bibAnnote}[2]{%
  \begin{quotation}\noindent\textsc{Key:} #1\\
  \textsc{Annotation:}\ #2\end{quotation}}
\providecommand{\eprint}[2][]{\url{#2}}

\bibitem{Pritchard2000b}
Pritchard JK, Stephens M, Rosenberg NA, Donnelly P (2000) {Association mapping
  in structured populations.}
\newblock American journal of human genetics 67: 170--81.
\bibAnnoteFile{Pritchard2000b}

\bibitem{Price2006}
Price AL, Patterson NJ, Plenge RM, Weinblatt ME, Shadick NA, et~al. (2006)
  {Principal components analysis corrects for stratification in genome-wide
  association studies}.
\newblock Nature Genetics 38: 904--909.
\bibAnnoteFile{Price2006}

\bibitem{reich2011denisova}
Reich D, Patterson N, Kircher M, DelÞn F, Nandineni MR, et~al. (2011) Denisova
  admixture and the first modern human dispersals into southeast asia and
  oceania.
\newblock Am J Hum Gen 89: 1--13.
\bibAnnoteFile{reich2011denisova}

\bibitem{moorjani2011history}
Moorjani P, Patterson N, Hirschhorn JN, Keinan A, Hao L, et~al. (2011) The
  history of african gene flow into southern europeans, levantines, and jews.
\newblock PLOS Genetics .
\bibAnnoteFile{moorjani2011history}

\bibitem{wang2013apparent}
Wang S, Lachance J, Tishkoff SA, Hey J, Xing J (2013) Apparent variation in
  neanderthal admixture among african populations is consistent with gene flow
  from non-african populations.
\newblock Genome Biol Evol 5: 2075--2081.
\bibAnnoteFile{wang2013apparent}

\bibitem{pritchard2000inference}
Pritchard JK, Stephens M, Donnelly P (2000) Inference of population structure
  using multilocus genotype data.
\newblock Genetics 155.
\bibAnnoteFile{pritchard2000inference}

\bibitem{Gilbert2012}
Gilbert KJ, Andrew RL, Bock DG, Franklin MT, Kane NC, et~al. (2012)
  Recommendations for utilizing and reporting population genetic analyses: the
  reproducibility of genetic clustering using the program structure.
\newblock Molecular ecology 21: 4925--4930.
\bibAnnoteFile{Gilbert2012}

\bibitem{Box1980}
Box GEP (1980) {Sampling and Bayes' inference in scientific modelling and
  robustness}.
\newblock Journal of the Royal Statistical Society Series A ( \ldots 143:
  383--430.
\bibAnnoteFile{Box1980}

\bibitem{Rubin1984}
Rubin D (1984) {Bayesianly Justifiable and Relevant Frequency Calculations for
  the Applied Statistician}.
\newblock The Annals of Statistics 12: 1151--1172.
\bibAnnoteFile{Rubin1984}

\bibitem{Meng1993}
MENG XL, RUBIN DB (1993) {Maximum likelihood estimation via the ECM algorithm:
  A general framework}.
\newblock Biometrika 80: 267--278.
\bibAnnoteFile{Meng1993}

\bibitem{gelman1996posterior}
Gelman A, Meng X, Stern H (1996) posterior predictive assessment of model
  fitness via realized discrepancies.
\newblock Statistica Sinica 6: 733--807.
\bibAnnoteFile{gelman1996posterior}

\bibitem{Gelman2013}
Gelman A, Shalizi CR (2013) {Philosophy and the practice of Bayesian
  statistics.}
\newblock The British journal of mathematical and statistical psychology 66:
  8--38.
\bibAnnoteFile{Gelman2013}

\bibitem{Rosenberg2002}
Rosenberg NA, Pritchard JK, Weber JL, Cann HM, Kidd KK, et~al. (2002) {Genetic
  Structure of Human Populations}.
\newblock Science 298: 2381--2385.
\bibAnnoteFile{Rosenberg2002}

\bibitem{Engelhardt2010}
Engelhardt BE, Stephens M (2010) {Analysis of Population Structure: A Unifying
  Framework and Novel Methods Based on Sparse Factor Analysis}.
\newblock PLoS Genet 6.
\bibAnnoteFile{Engelhardt2010}

\bibitem{Nelson2008}
Nelson MR, Bryc K, King KS, Indap A, Boyko AR, et~al. (2008) {The Population
  Reference Sample, POPRES: A Resource for Population, Disease, and
  Pharmacological Genetics Research}.
\newblock American Journal of Human Genetics 83: 347--358.
\bibAnnoteFile{Nelson2008}

\bibitem{1000genomes}
Durbin RM, Altshuler DL, Abecasis GR, Bentley DR, Chakravarti A, et~al. (2010)
  {A map of human genome variation from population-scale sequencing}.
\newblock Nature 467: 1061--1073.
\bibAnnoteFile{1000genomes}

\bibitem{Reich2009}
Reich D, Thangaraj K, Patterson N, Price AL, Singh L (2009) {Reconstructing
  Indian population history}.
\newblock Nature 461: 489--494.
\bibAnnoteFile{Reich2009}

\bibitem{Gelman2004}
Gelman A (2004) {Exploratory Data Analysis for Complex Models}.
\newblock Journal of Computational and Graphical Statistics 13: 755--779.
\bibAnnoteFile{Gelman2004}

\bibitem{Altshuler2010}
Altshuler DM, Gibbs RA, Peltonen L, Dermitzakis E, Schaffner SF, et~al. (2010)
  {Integrating common and rare genetic variation in diverse human populations.}
\newblock Nature 467: 52--8.
\bibAnnoteFile{Altshuler2010}

\bibitem{Rosenberg2003}
Rosenberg NA, Li LM, Ward R, Pritchard JK (2003) {Informativeness of genetic
  markers for inference of ancestry.}
\newblock American journal of human genetics 73: 1402--22.
\bibAnnoteFile{Rosenberg2003}

\bibitem{Novembre2008}
Novembre J, Johnson T, Bryc K, Kutalik Z, Boyko AR, et~al. (2008) {Genes mirror
  geography within Europe}.
\newblock Nature 456: 98--101.
\bibAnnoteFile{Novembre2008}

\bibitem{Bishop1990}
Bishop DT, Williamson JA (1990) {The power of identity-by-state methods for
  linkage analysis.}
\newblock American journal of human genetics 46: 254--65.
\bibAnnoteFile{Bishop1990}

\bibitem{Weir1984}
Weir BS, Cockerham CC (1984) {Estimating F-Statistics for the Analysis of
  Population Structure} : 1358----1370.
\bibAnnoteFile{Weir1984}

\bibitem{Stephens1994}
Stephens JC, Briscoe D, O'Brien SJ (1994) {Mapping by admixture linkage
  disequilibrium in human populations: limits and guidelines.}
\newblock American journal of human genetics 55: 809--24.
\bibAnnoteFile{Stephens1994}

\bibitem{Conrad2006}
Conrad DF, Jakobsson M, Coop G, Wen X, Wall JD, et~al. (2006) {A worldwide
  survey of haplotype variation and linkage disequilibrium in the human
  genome}.
\newblock Nature Genetics 38: 1251--1260.
\bibAnnoteFile{Conrad2006}

\bibitem{Excoffier2013}
Excoffier L, Dupanloup I, Huerta-S\'{a}nchez E, Sousa VC, Foll M (2013) {Robust
  demographic inference from genomic and SNP data.}
\newblock PLoS genetics 9: e1003905.
\bibAnnoteFile{Excoffier2013}

\bibitem{Daly2001}
Daly MJ, Rioux JD, Schaffner SF, Hudson TJ, Lander ES (2001) {High-resolution
  haplotype structure in the human genome.}
\newblock Nature genetics 29: 229--32.
\bibAnnoteFile{Daly2001}

\bibitem{Gabriel2002}
Gabriel SB, Schaffner SF, Nguyen H, Moore JM, Roy J, et~al. (2002) {The
  structure of haplotype blocks in the human genome.}
\newblock Science (New York, NY) 296: 2225--9.
\bibAnnoteFile{Gabriel2002}

\bibitem{Greenwood2004}
Greenwood TA, Rana BK, Schork NJ (2004) {Human Haplotype Block Sizes Are
  Negatively Correlated With Recombination Rates} : 1358--1361.
\bibAnnoteFile{Greenwood2004}

\bibitem{Falush2003}
Falush D, Stephens M, Pritchard JK (2003) {Inference of population structure
  using multilocus genotype data: linked loci and correlated allele
  frequencies.}
\newblock Genetics 164: 1567--1587.
\bibAnnoteFile{Falush2003}

\bibitem{Fledel2011}
Fledel-Alon A, Leffler EM, Guan Y, Stephens M, Coop G, et~al. (2011) {Variation
  in human recombination rates and its genetic determinants.}
\newblock PloS one 6: e20321.
\bibAnnoteFile{Fledel2011}

\bibitem{Slatkin2008}
Slatkin M (2008) {Linkage disequilibrium--understanding the evolutionary past
  and mapping the medical future.}
\newblock Nature reviews Genetics 9: 477--85.
\bibAnnoteFile{Slatkin2008}

\bibitem{Cover1991}
Cover TM, Thomas JA (1991) {Elements of Information Theory (Wiley Series in
  Telecommunications and Signal Processing)}.
\newblock Wiley-Interscience, 576 pp.
\newblock
  \urlprefix\url{http://www.amazon.com/Elements-Information-Theory-Telecommunications-Processing/dp/0471062596}.
\bibAnnoteFile{Cover1991}

\bibitem{Altshuler2005}
{The International HapMap Consortium} (2005) {A haplotype map of the human
  genome.}
\newblock Nature 437: 1299--320.
\bibAnnoteFile{Altshuler2005}

\bibitem{Shifman2003}
Shifman S (2003) {Linkage disequilibrium patterns of the human genome across
  populations}.
\newblock Human Molecular Genetics 12: 771--776.
\bibAnnoteFile{Shifman2003}

\bibitem{McEvoy2011}
McEvoy BP, Powell JE, Goddard ME, Visscher PM (2011) {Human population
  dispersal "Out of Africa" estimated from linkage disequilibrium and allele
  frequencies of SNPs.}
\newblock Genome research 21: 821--9.
\bibAnnoteFile{McEvoy2011}

\bibitem{WTCCC}
 (2007) {Genome-wide association study of 14,000 cases of seven common diseases
  and 3,000 shared controls.}
\newblock Nature 447: 661--78.
\bibAnnoteFile{WTCCC}

\bibitem{Hoggart2003}
Hoggart CJ, Parra EJ, Shriver MD, Bonilla C, Kittles RA, et~al. (2003) {Control
  of confounding of genetic associations in stratified populations.}
\newblock American journal of human genetics 72: 1492--1504.
\bibAnnoteFile{Hoggart2003}

\bibitem{Satten2001}
Satten GA, Flanders WD, Yang Q (2001) {Accounting for unmeasured population
  substructure in case-control studies of genetic association using a novel
  latent-class model.}
\newblock American journal of human genetics 68: 466--77.
\bibAnnoteFile{Satten2001}

\bibitem{Devlin1999}
Devlin B, Roeder K (1999) {Genomic control for association studies.}
\newblock Biometrics 55: 997--1004.
\bibAnnoteFile{Devlin1999}

\bibitem{Patterson2004}
Patterson N, Hattangadi N, Lane B, Lohmueller KE, Hafler DA, et~al. (2004)
  {Methods for high-density admixture mapping of disease genes.}
\newblock American journal of human genetics 74: 979--1000.
\bibAnnoteFile{Patterson2004}

\bibitem{Marchini2004}
Marchini J, Cardon LR, Phillips MS, Donnelly P (2004) {The effects of human
  population structure on large genetic association studies.}
\newblock Nature genetics 36: 512--7.
\bibAnnoteFile{Marchini2004}

\bibitem{Patterson2006}
Patterson N, Price AL, Reich D (2006) Population structure and eigenanalysis.
\newblock PLOS genetics 2: e190.
\bibAnnoteFile{Patterson2006}

\bibitem{Price2010}
Price AL, Zaitlen NA, Reich D, Patterson N (2010) {New approaches to population
  stratification in genome-wide association studies.}
\newblock Nature reviews Genetics 11: 459--63.
\bibAnnoteFile{Price2010}

\bibitem{Jorde2004}
Jorde LB, Wooding SP (2004) {Genetic variation, classification and 'race'}.
\newblock Nature Genetics .
\bibAnnoteFile{Jorde2004}

\bibitem{Serre2004}
Serre D, P\"{a}\"{a}bo S (2004) {Evidence for gradients of human genetic
  diversity within and among continents.}
\newblock Genome research 14: 1679--85.
\bibAnnoteFile{Serre2004}

\bibitem{Tang2006}
Tang H, Coram M, Wang P, Zhu X, Risch N (2006) {Reconstructing genetic ancestry
  blocks in admixed individuals.}
\newblock American journal of human genetics 79: 1--12.
\bibAnnoteFile{Tang2006}

\bibitem{Sankararaman2008}
Sankararaman S, Kimmel G, Halperin E, Jordan MI (2008) {On the inference of
  ancestries in admixed populations.}
\newblock Genome research 18: 668--75.
\bibAnnoteFile{Sankararaman2008}

\bibitem{Evanno2005}
Evanno G, Regnaut S, Goudet J (2005) {Detecting the number of clusters of
  individuals using the software STRUCTURE: a simulation study.}
\newblock Molecular ecology 14: 2611--20.
\bibAnnoteFile{Evanno2005}

\bibitem{Huelsenbeck2007}
Huelsenbeck JP, Andolfatto P (2007) {Inference of population structure under a
  Dirichlet process model.}
\newblock Genetics 175: 1787--802.
\bibAnnoteFile{Huelsenbeck2007}

\bibitem{Teh2006}
Teh YW, Jordan MI, Beal MJ, Blei DM (2003) {Hierarchical Dirichlet Processes}.
\newblock Journal of the American Statistical Association 101.
\bibAnnoteFile{Teh2006}

\bibitem{Blei2003b}
Blei D, Griffiths T, Jordan M, Tenenbaum J (2003) {Hierarchical Topic Models
  and the Nested Chinese Restaurant Process.}
\newblock NIPS .
\bibAnnoteFile{Blei2003b}

\bibitem{Adams2010}
Adams R, Ghahramani Z, Jordan M (2010) {Tree-Structured Stick Breaking for
  Hierarchical Data.}
\newblock NIPS : 1--9.
\bibAnnoteFile{Adams2010}

\bibitem{Ardlie2002}
Ardlie KG, Lunetta KL, Seielstad M (2002) {Testing for population subdivision
  and association in four case-control studies.}
\newblock American journal of human genetics 71: 304--11.
\bibAnnoteFile{Ardlie2002}

\bibitem{Pickrell2012}
Pickrell JK, Pritchard JK (2012) {Inference of population splits and mixtures
  from genome-wide allele frequency data.}
\newblock PLoS genetics 8: e1002967.
\bibAnnoteFile{Pickrell2012}

\bibitem{Gautier2013}
Gautier M, Vitalis R (2013) {Inferring population histories using genome-wide
  allele frequency data.}
\newblock Molecular biology and evolution 30: 654--68.
\bibAnnoteFile{Gautier2013}

\bibitem{Nelson2012}
Nelson MR, Wegmann D, Ehm MG, Kessner D, {St Jean} P, et~al. (2012) {An
  abundance of rare functional variants in 202 drug target genes sequenced in
  14,002 people.}
\newblock Science (New York, NY) 337: 100--4.
\bibAnnoteFile{Nelson2012}

\bibitem{Lee1999}
D LD, Seung HS (1999) {Learning the parts of objects by non-negative matrix
  factorization}.
\newblock Nature 401: 788--791.
\bibAnnote{Lee1999}{A specific type of generalized factor analysis.
  Non-negative matrix factorization paper.}

\bibitem{Rasmussen2000}
Rasmussen CE (2000) {The Infinite Gaussian Mixture Model} .
\bibAnnoteFile{Rasmussen2000}

\bibitem{Yang2012}
Yang WY, Novembre J, Eskin E, Halperin E (2012) {A model-based approach for
  analysis of spatial structure in genetic data}.
\newblock Nature Genetics 44: 725--731.
\bibAnnoteFile{Yang2012}

\bibitem{Paisley2012}
Paisley J, Wang C, Blei DM (2012) {The Discrete Infinite Logistic Normal
  Distribution}.
\newblock Bayesian Analysis 7: 997--1034.
\bibAnnoteFile{Paisley2012}

\bibitem{Kang2010}
Kang HM, Sul JH, Service SK, Zaitlen Na, Kong SY, et~al. (2010) {Variance
  component model to account for sample structure in genome-wide association
  studies.}
\newblock Nature genetics 42: 348--54.
\bibAnnoteFile{Kang2010}

\bibitem{Listgarten2012}
Listgarten J, Lippert C, Kadie CM, Davidson RI, Eskin E, et~al. (2012)
  {Improved linear mixed models for genome-wide association studies.}
\newblock Nature methods 9: 525--6.
\bibAnnoteFile{Listgarten2012}

\bibitem{Zhou2012}
Zhou X, Stephens M (2012) {Genome-wide efficient mixed-model analysis for
  association studies.}
\newblock Nature genetics 44: 821--4.
\bibAnnoteFile{Zhou2012}

\bibitem{Runcie2013}
Runcie DE, Mukherjee S (2013) {Dissecting High-Dimensional Phenotypes with
  Bayesian Sparse Factor Analysis of Genetic Covariance Matrices.}
\newblock Genetics : genetics.113.151217--.
\bibAnnoteFile{Runcie2013}

\bibitem{alexander2009fast}
David H~Alexander JN, Lange K (2009) Fast model-based estimation of ancestry in
  unrelated individuals.
\newblock Genome Res .
\bibAnnoteFile{alexander2009fast}

\bibitem{Dempster1977}
Dempster A, Laird N, Rubin DB (1977) {Maximum likelihood from incomplete data
  via the EM algorithm}.
\newblock Journal of the Royal Statistical Society, Series B 39: 1--38.
\bibAnnoteFile{Dempster1977}

\bibitem{wright1969evolution}
Wright S (1969) Evolution and the Genetics of Populations, Vol. II. The Theory
  of Gene Frequencies.
\newblock University of Chicago Press.
\bibAnnoteFile{wright1969evolution}

\bibitem{Stephens2009}
Stephens M, Balding DJ (2009) {Bayesian statistical methods for genetic
  association studies.}
\newblock Nature reviews Genetics 10: 681--90.
\bibAnnoteFile{Stephens2009}

\end{thebibliography}

\end{document}